\documentclass[AMA,STIX1COL, doublespace]{WileyNJD-v2}
\usepackage{moreverb}

\usepackage{xcolor}
\usepackage{graphicx}
\usepackage{bm}
\usepackage{amsmath}
\usepackage{caption}
\usepackage{subcaption}
\usepackage{ulem}

\usepackage{rotating}
\usepackage{fancyhdr}
\usepackage{environ}
\usepackage{pgf}
\usepackage{tikz}
\usetikzlibrary{arrows,automata}
\usetikzlibrary{positioning}
\tikzset{
    state/.style={
           rectangle,
           rounded corners,
           draw=black, very thick,
           minimum height=2em,
           inner sep=2pt,
           text centered,
           },
}
\tikzset{
    mycomment/.style={
           rectangle,
           rounded corners,
           draw={rgb:black,1;white,2},
           minimum height=2em,
           inner sep=2pt,
           text centered,
           text=white,
           fill={rgb:black,1;white,2},
           },
}
\NewEnviron{myequation}{%
    \begin{equation}
    \scalebox{0.8}{$\BODY$}
    \end{equation}
    }
    
    \NewEnviron{mymath}{%
   $ \scalebox{0.8}{$\BODY$}$
    }
    
        \NewEnviron{cmtmath}{%
   $ \scalebox{0.5}{$\BODY$}$
    }

\newcommand{\package}[1]{{\normalfont\fontseries{b}\selectfont #1}} 

\newcommand{\ba}    {\begin{array}}
	\newcommand{\ea}    {\end{array}}
\newcommand{\be}    {\begin{equation}}
	\newcommand{\ee}    {\end{equation}}
\newcommand{\bea}    {\begin{eqnarray}}
	\newcommand{\eea}    {\end{eqnarray}}
\newcommand{\nn}     {\nonumber}  

\newcommand{\uR}       {\mbox{\boldmath$R$}}

\newcommand{\uX}       {\mbox{\boldmath$X$}}
\newcommand{\ux}{\mathbf{x}}
\newcommand{\uY}       {\mbox{\boldmath$Y$}}

\newcommand{\ualpha}            {\mbox{\boldmath$\alpha$}}
\newcommand{\ubeta}             {\mbox{\boldmath$\beta$}}
\newcommand{\ugamma}            {\mbox{\boldmath$\gamma$}}

\newcommand{\uiota}             {\mbox{\boldmath$\uiota$}}

\newcommand\BibTeX{{\rmfamily B\kern-.05em \textsc{i\kern-.025em b}\kern-.08em
T\kern-.1667em\lower.7ex\hbox{E}\kern-.125emX}}

\articletype{Article Type}%

\received{<day> <Month>, <year>}
\revised{<day> <Month>, <year>}
\accepted{<day> <Month>, <year>}

\begin{document}

\title{Cumulative Logit Ordinal Regression with Proportional Odds under Nonignorable Missing Responses – Application to Phase III Trial}

\author[1]{Arnab Kumar Maity}

\author[2]{Huaming Tan}

\author[3]{Vivek Pradhan}

\author[4]{Soutir Bandyopadhyay}

\authormark{Arnab Kumar Maity \textsc{et al}}

\address[1]{\orgname{Boehringer Ingelheim Pharmaceuticals, Inc.}, \orgaddress{\state{Ridgefield, Connecticut}, \country{USA}}}

\address[2]{\orgname{Pfizer, Inc.}, \orgaddress{\state{Groton, Connecticut}, \country{USA}}}

\address[3]{\orgname{Pfizer, Inc.},  \orgaddress{\state{Cambridge, Massachusetts}, \country{USA}}}

\address[4]{\orgdiv{Department of Applied Mathematics and Statistics}, \orgname{Colorado School of Mines}, \orgaddress{\state{Golden, Colorado}, \country{USA}}}

\corres{*Arnab Kumar Maity, Boehringer Ingelheim, California, USA. \email{arnab.maity@boehringer-ingelheim.com}}

\presentaddress{California, USA}

\abstract[Abstract]{Missing data are inevitable in clinical trials, and trials that produce categorical ordinal responses are not exempted from this. Typically, missing values in the data occur due to different missing mechanisms, such as missing completely at random, missing at random, and missing not at random. Under a specific missing data regime, when the conditional distribution of the missing data is dependent on the ordinal response variable itself along with other predictor variables, then the missing data mechanism is called \textit{nonignorable}. In this article we propose an expectation maximization based algorithm for fitting a proportional odds regression model when the missing responses are nonignorable. We report results from an extensive simulation study to illustrate the methodology and its finite sample properties. We also apply the proposed method to a recently completed Phase III psoriasis study using an investigational compound. The corresponding SAS program is provided.}

\keywords{Cumulative Logit; EM Algorithm; Nonignorbale Missing; Ordinal Response; Proportional Odds; Randomized Trial.}


\maketitle


\section{Introduction}  \label{sec:intro}


In clinical research, a fundamental objective is to assess the efficacy of a drug or treatment in alleviating a disease compared to a placebo or control group. However, a common challenge in clinical trials is the presence of missing data, which can arise for various reasons, including participant dropout or incomplete data collection. When the outcome of interest is ordinal, such as a graded scale of disease severity, the impact of missing data becomes even more pronounced. If these missing observations are not adequately taken into account, the resulting analysis yield biased estimates, leading to potentially misleading conclusions about the effectiveness of the treatment. Hence, addressing the issue of missing data is critical for ensuring valid and reliable inferences. In this article, we address the problem of missing data in trials with ordinal outcomes and propose a robust methodology to ensure more accurate assessment of treatment efficacy.

Our discussion will be put in the context of the analysis of the data of a double-blind randomized Phase III clinical trial (\href{https://clinicaltrials.gov/ct2/show/NCT01309737}{clinicaltrials.gov: NCT01309737}; Papp et al. \cite{papp2015tofacitinib}), evaluating a compound for the treatment of psoriasis. This study enrolled $383$, $381$, and $196$ participants in two active dose groups (5 mg and 10 mg) and a placebo group, respectively. One of the primary endpoints was the Physician’s Global Assessment (PGA) score, which used a five-point scale (0, 1, 2, 3, 4), with $0$ indicating clear skin (best outcome) and $4$ indicating severe disease (worst outcome). Participants were classified as responders if their PGA score at Week-16 was either $0$ or $1$. The primary clinical objective was to estimate the odds of achieving a PGA score of $ \le 1$ for each dose group compared to placebo, adjusting for covariates. A secondary aim was to determine whether any additional predictors were statistically and clinically relevant in explaining the PGA score.

During the analysis, it was observed that 13.5\% of participants did not have a recorded PGA score at the landmark time point of week 16. As a result, the statistical inferences related to the PGA endpoint were suspected to be biased due to the missing data, prompting the need for further sensitivity analyses. This challenge inspired the development of a methodology to enable valid statistical inference in the presence of missing data. 

Missing data mechanisms can broadly be classified into three categories -- missing completely at random (MCAR), missing at random (MAR), and missing not at random (MNAR) \citep{little2019statistical}. The missing data mechanism is MCAR when the probability of occurring the missing data is not dependent in any of the data points. The mechanism is referred to as MAR if the probability of missing data is function of other variables in the dataset. Under MNAR and under a regression set up, if the conditional distribution of the missing data is dependent on the \textcolor {black} {unobserved} response variable, then the missing data mechanism is nonignorable. This article centers around the theory and methodology related to noningnorable missing data mechanism. 

The literature on ordinal models is extensive and beyond the scope of this article to comprehensively review. McKinley et al. \cite{mckinley2015bayesian} discuss Bayesian advancements, while Agresti \cite{agresti2003categorical} provides a thorough overview of the frequentist framework, including proportional odds models. Significant work has also been conducted on inference procedures for missing data under missing completely at random (MCAR) and missing at random (MAR) assumptions in categorical and ordinal data analysis \citep{chen2020testing, wu2015comparison, jia2019evaluating}. These methods generally rely on the assumption that the missingness mechanism is ignorable, meaning the likelihood function can be correctly specified without explicitly modeling the missing data mechanism. Within this context, Kenward et al. \cite{kenward1994application} study proportional odds models under the MAR assumption, which, under standard conditions, is treated as ignorable for likelihood-based inference.

Despite these advances, there remains a notable gap in the literature addressing nonignorable missing data mechanisms in ordinal response models, where the missingness depends on the unobserved responses themselves. In this article, we propose a parameter estimation technique based on the expectation-maximization (EM) algorithm, specifically designed for settings where responses are missing and the missingness mechanism is nonignorable. This approach builds on the general framework introduced by Ibrahim and Lipsitz \cite{ibrahim1996parameter}, who developed an EM algorithm for parameter estimation in binomial regression models with nonignorable missing responses. Subsequent extensions of this framework \citep{maity2019bias} have demonstrated strong performance in simulation studies, providing a solid foundation for its adaptation to ordinal regression models in this study.


The remainder of this article is organized as follows: Section \ref{sec:proplikelihood} outlines the general framework of the likelihood. Section \ref{section_method} introduces the proposed method, which builds upon the work of Ibrahim and Lipsitz \cite{ibrahim1996parameter}, extending their methodology to the proportional odds model to describe the relationship between responses and covariates, while incorporating a binary logistic regression model to account for the missing data mechanism. In Section \ref{section_simulation}, we present simulation studies to evaluate the performance of the proposed method, focusing on bias, mean squared error (MSE), and 95\% coverage probability. Section \ref{section_real_data} demonstrates the application of the method using data from a psoriasis study. Finally, Section \ref{section_discussion} provides concluding remarks.

\section{The Likelihood of Proportional Odds Model} \label{sec:proplikelihood}

Without loss of generality, suppose the random variable $Y_i$, for $i = 1, 2, \ldots, n$ observations, takes one of the ordered discrete values $1, 2, \ldots, J$. Let the probability that the response of individual $i$ falls into category $j$ be denoted by $\pi_{ij} \equiv P(Y_i = j | X_i)$, where $X_i$ is the $i$-th row of the regressor matrix $\uX$, which has \textcolor {black} {dimension} $n \times p$ and includes $p$ regressors. For the \textcolor {black} {cumulative logit proportional odds model}, the relationship between $\pi_{ij}$ and $X_i$ can be expressed as:

\bea  \label{equation_PO_regression}
\log\left(\frac{P(Y_i \le j)}{1- P(Y_i \le j)}\right)&=& \beta_{0j} + {X_i}^{T}\ubeta,  \quad j=1, 2, \ldots, J - 1,
\eea
where $ P(Y_i \le j) = \sum_{l = 1}^j \pi_{il} $, $\beta_{0j}$ represent the intercept term for each category $j$ of the ordered discrete values, and $\ubeta \equiv (\beta_1, \beta_2, \ldots, \beta_p)^T$.

Let $f_i(Y_i | \beta_{0j}, \ubeta, X_i)$ represent the probability of observing response $Y_i$ for the $i$-th observation, given the covariates $X_i$ and the parameters $\beta_{0j}$ and $\ubeta$. The joint probability for all $n$ observations, often referred to as the likelihood of the observed data, is given by:

\bea
L(\ubeta_0, \ubeta | \uX, \uY) &=& \prod_{i=1}^{n} f_i(Y_i | \ubeta_0, \ubeta, X_i), \nn \\
&=& \prod_{i=1}^{n}\prod_{j=1}^{J} \pi_{ij} (\ubeta_{0j}, \ubeta)^{y_{ij}}
\eea
where $\ubeta_0 = (\beta_{01}, \ldots, \beta_{0(J-1)})^T$, and $y_{ij}$ is the binary indicator of the responses for the $i$-th observation. 




Let $ \uR $ be the missing indicator 
vector whose $i$-th element is defined as,
\bea  \label{equation_missing_data_indicator}
R_i = \left\{ 
\begin{array}{l l}
	1 \quad \text{if $Y_i$ is missing}\\
	0 \quad \text{if $Y_i$ is observed},  \nn \\
\end{array} \right.
\eea

The missing data mechanism can be modeled as:
\bea \label{equation_missing_model}
p_i(R_i | \ualpha, Z_i) = P(R_i = 1 | \ualpha, Z_i) = \frac{\exp(Z_i^T \ualpha)}{1 + \exp(Z_i^T \ualpha)}, \quad i = 1, 2, \ldots, n,
\eea
where $Z_i = (X_i^T, Y_i)^T$ and $\ualpha = (\alpha_0, \alpha_1, \ldots, \alpha_{p+1})^T$ is a parameter vector of dimension $p+2$. When $\ualpha$ is a null vector, the missing data mechanism is missing completely at random (MCAR). Under the missing at random (MAR) assumption, if $\alpha_{p+1} = 0$, the missingness mechanism is ignorable. However, when $\alpha_{p+1} \neq 0$, the missingness mechanism becomes nonignorable (MNAR).

Let $\ugamma = (\ubeta_0^T, \ubeta^T, \ualpha^T)^T$ denote the complete parameter vector. The joint log-likelihood of the observed data can be expressed as:
\bea
l(\ugamma | \uX, \uY, \uR) &=& \log\left\{\prod_{i=1}^{n} f_i(Y_i | \ubeta_0, \ubeta, X_i) p_i(R_i | \ualpha, Z_i)\right\} \nn \\
&=& \sum_{i=1}^{n} \left\{\log(f_i(Y_i | \ubeta_0, \ubeta, X_i)) + \log(p_i(R_i | \ualpha, Z_i))\right\}.
\eea

For simplicity, we redefine $\ubeta^T \equiv (\ubeta_0^T, \ubeta^T)$. Thus, the joint log-likelihood becomes:
\bea
l(\ugamma | \uX, \uY, \uR) &=& \sum_{i=1}^{n} \left\{\log(f_i(Y_i | \ubeta, X_i)) + \log(p_i(R_i | \ualpha, Z_i))\right\}.
\eea


\section{The EM Algorithm}  \label{section_method}
Ibrahim and Lipsitz \cite{ibrahim1996parameter} proposed an Expectation-Maximization (EM) algorithm for estimating parameters in binary regression models with nonignorable missing responses. Building on their framework, we extend the EM algorithm to accommodate ordinal response variables. This approach provides a robust solution for handling nonignorable missing data in the context of cumulative logit proportional odds models.
\subsection{The E Step}  \label{section_weights}
The EM algorithm involves maximizing the expected complete-data log-likelihood, where the expectation is taken over the missing data conditioned on the observed data and current parameter estimates. Specifically, for the $i$-th individual, the expected log-likelihood can be expressed as:

\[
E[l(\ugamma|X_i, Y_i, R_i)] =
\begin{cases}
\sum_{j = 1}^{J} l(\ugamma|X_i, Y_i, R_i) f_i(Y_i| X_i, R_i, \ugamma), & \text{if $Y_i$ is missing}, \\
l(\ugamma|X_i, Y_i, R_i), & \text{if $Y_i$ is observed}.
\end{cases}
\]
where $l(\ugamma|X_i, Y_i, R_i)$ is the complete-data log-likelihood, and $p(Y_i| X_i,R_i, \ugamma)$ is the conditional probability of missing data given the observed data. These probabilities can be interpreted as weights, denoted by $w_{iy_i}$, and can be expressed as the joint distribution of $(Y_i, R_i| X_i, \ugamma)$ as:

\[
w_{iy_i} =
\begin{cases}
\frac{f_i(Y_i, R_i| X_i, \ugamma)}{\sum_{j = 1}^{J} f_i(Y_i, R_i| X_i, \ugamma)}, & \text{if $Y_i$ is missing}, \\
1, & \text{if $Y_i$ is observed}.
\end{cases}
\]
This can be simplified further as:
\[
w_{iy_i} =
\begin{cases}
\frac{f_i(Y_i | X_i, \ubeta) p_i(R_i | Z_i, \ualpha)}{\sum_{j = 1}^{J} f_i(Y_i | X_i, \ubeta) p_i(R_i | Z_i, \ualpha)}, & \text{if $Y_i$ is missing}, \\
1, & \text{if $Y_i$ is observed}.
\end{cases}
\]

For the (t+1)-th iteration, the expected complete-data log-likelihood for all $n$ observations is updated as:
\[
Q(\ugamma|\ugamma^{(t)}) = \sum_{i=1}^{n} \sum_{j = 1}^{J} w_{iy_i}^{(t)} l(\ugamma| X_i, Y_i, Z_i, R_i).
\]
Expanding this further:
\[
Q(\ugamma|\ugamma^{(t)}) = \sum_{i=1}^{n} \sum_{j = 1}^{J} w_{iy_i}^{(t)} \bigl\{ l(\ubeta|X_i, Y_i) + l(\ualpha|Z_i, R_i)\bigr\}.
\]
Simplifying in terms of component functions:
\[
Q(\ugamma|\ugamma^{(t)}) = \sum_{i=1}^{n} \sum_{j = 1}^{J} w_{iy_i}^{(t)} \bigl\{ Q_{1}(\ubeta|\ugamma^{(t)}) + Q_{2}(\ualpha|\ugamma^{(t)}) \bigr\}.
\]
Here, $\ugamma^{(t)}$ is the parameter vector at the $t$-th iteration. When $Y_i$ is observed, the summation over all possible values of $Y_i$ is unnecessary, as the observed value directly contributes to the calculation, bypassing the need for imputation or expectation over missing values.

\subsection{The M Step}  \label{section_M_step}

The maximization of the M-step involves two separate maximizations -- $ Q_1(\bm{\beta}|\bm{\gamma}^{(t)}) $ and $  Q_2(\bm{\alpha}|\bm{\gamma}^{(t)}) $ with the updated weight $ w_{i y_i}^{(t)}$ at $(t+1)$-th stage. This can be done by fitting two separate models, one with a proportional odds model and the other with binary logistic regression models with specified updated weights.  The maximization steps follow a route similar to that given in Ibrahim and Lipsitz \cite{ibrahim1996parameter}. Letting $ \dot{Q} $ and $ \ddot{Q} $ denote the matrix of first derivatives and second derivatives of $ Q(\gamma|\gamma^{(t)})$, we define 
\begin{align*}
	\dot{Q}(\bm{\gamma}|\bm{\gamma}^{(t)}) &\equiv \sum_{i = 1}^n \dot{q}(\bm{\gamma}|\bm{\gamma}^{(t)})  \\
	&= 
	\sum_{i=1}^{n} \sum_{j = 1}^{J} w_{i y_i}^{(t)} \frac{\partial l(\bm{\bm{\gamma}}| \ux_i, y_i, r_i)}{\partial \bm{\gamma}}
\end{align*}
By the virtue of construction, 
\[
\dot{Q}(\bm{\gamma}|\bm{\gamma}^{(t)}) = \begin{pmatrix}
	\dot{Q}_1(\bm{\beta}|\bm{\beta}^{(t)}) \\
	\dot{Q}_1(\bm{\alpha}|\bm{\alpha}^{(t)})
\end{pmatrix},
\]
where $ \dot{Q}_1(\bm{\beta}|\bm{\beta}^{(t)}) $ and $ \dot{Q}_2(\bm{\alpha}|\bm{\alpha}^{(t)}) $ have $ (J-1+p) $ and $ (p + 2) $ components. The $r$-th components are given by
\[
\dot{Q}_{1r}(\bm{\beta}|\bm{\beta}^{(t)}) = \sum_{i = 1}^n \sum_{j = 1}^{J} w_{iy_i}^{(t)} P(Y_i \le j)(1-P(Y_i \le j)\left\{\frac{Y_{ij}}{\pi_{ij}} - \frac{Y_{i(j+1)}}{\pi_{i(j+1)}}\right\}\kappa_{ijr}, \quad r = 1, \ldots, (J - 1 + p),
\]
where $\kappa_{ijr}$ is the $(j, r)$-th component of the $(J-1) \times (J-1+p)$ matrix
\[
\kappa_i=\begin{pmatrix}
    1&0&\dots&0&-X_i^{T} \\
    0&1&\dots&0&-X_i^{T} \\
    .&\dots& &\dots &\dots\\
    0&0&\dots&1&-X_i^{T} \\
\end{pmatrix}
\]
for $i=1, 2, \dots, n$ (see Kosmidis \cite{kosmidis2014improved})\\
and
\[
\dot{Q}_{2}(\bm{\alpha}|\bm{\alpha}^{(t)}) = \sum_{i = 1}^n \sum_{j = 1}^{J}w_{iy_i}^{(t)}(R_i - p_i)Z_{i},
\]
respectively. One would also require to compute 
\[
\ddot{Q}(\bm{\gamma}|\bm{\gamma}^{(t)}) = \begin{pmatrix}
	\ddot{Q}_1(\bm{\beta}|\bm{\beta}^{(t)}) & 0 \\
	0 & \ddot{Q}_1(\bm{\alpha}|\bm{\alpha}^{(t)})
\end{pmatrix}.
\]

The actual maximization during the M-step can be performed using numerical methods such as Newton-Raphson. Let $\bm{\gamma}^{(s,t)}$ denote the parameter estimate at the $s$-th iteration of the Newton-Raphson algorithm within the $t$-th iteration of the EM procedure. Similarly, let $\bm{\gamma}^{(t)}$ represent the final parameter estimate after convergence of the Newton-Raphson method within the $t$-th EM iteration. The parameter update at the $(s+1)$-th Newton-Raphson iteration within the $t$-th EM iteration is given by:

\[
\bm{\gamma}^{(s + 1, t)} = \bm{\gamma}^{(s, t)} - [\ddot{Q}(\bm{\gamma}^{(s, t)}|\bm{\gamma}^{(t)})]^{-1} \dot{Q}(\bm{\gamma}^{(s, t)}|\bm{\gamma}^{(t)}),
\]

where $\dot{Q}$ and $\ddot{Q}$ are the first and second derivatives of the $Q$-function, respectively.

The maximization of the weighted likelihood can be implemented using standard software after augmenting the missing data with all possible realizations of the response variable. For $Q_1$, representing the proportional odds model component, available software tools include \texttt{polr} from the \package{MASS} package, \texttt{vglm} from the \package{VGAM} package, or the \package{brglm2} package in \textsf{R}, as well as \texttt{PROC LOGISTIC}, \texttt{PROC CATMOD}, or \texttt{PROC GENMOD} in SAS. For $Q_2$, the binary logistic regression component, implementations such as \texttt{glm} in \textsf{R} or \texttt{PROC LOGISTIC} in SAS can be utilized. These routines enable efficient implementation of the M-step by leveraging standard maximum likelihood estimation techniques.

\subsection{The EM Algorithm}  \label{section_EM_algorithm}

\begin{figure}[h]
	\centering
	\includegraphics[height = 4 cm, width = 10 cm]{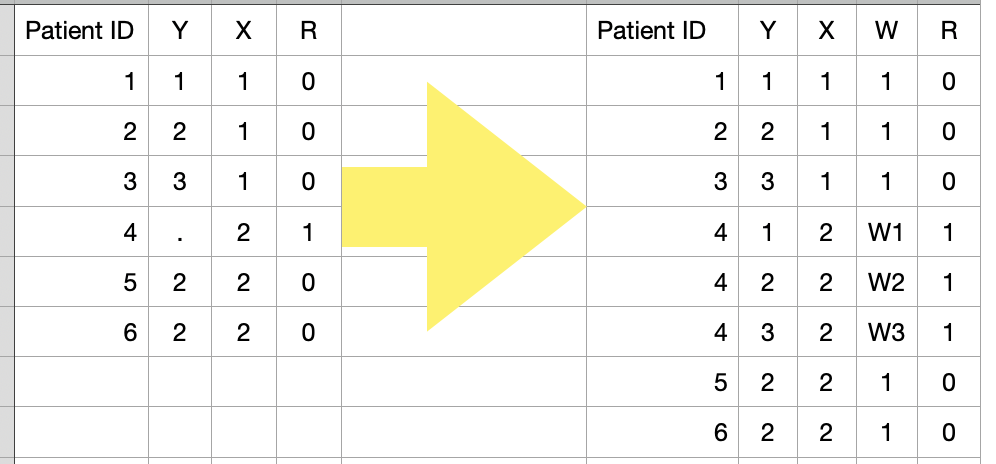}
	\caption{A schematic figure how to augment the data and add weights described in Section \ref{section_EM_algorithm}. \label{figure_data_augmentation}}
\end{figure}

We summarize all the steps of the EM algorithm here:
\begin{enumerate}
	\item Set the $ \bm{\gamma} $ with some arbitrary values.
	\item For each observed responses enter weight $w_{i y_i}$ as 1.
	\item For each missing response, augment all possible ordinal responses that would have been realized while keeping the same observed covariates. 
    \item Compute weight $w_{i y_i}$ for each augmented response as discussed in Section \ref{section_weights})  (see Figure \ref{figure_data_augmentation}). 
	\item Compose the E-step.
	\item M-step: Carry out the maximization using standard software that allows incorporating weight vector.
	\item Update weights $w_{i y_i}$ using the estimates obtained after M-step.
	\item Iterate Steps 3–7 until convergence.
\end{enumerate}

\subsection{Standard Error and Confidence Interval}

Louis \cite{louis1982finding} derived the standard deviation computation techniques when EM is used to estimate the parameters. Ibrahim and Lipsitz \cite{ibrahim1996parameter} followed this method and we state the similar proposal in the setting of our study. Note that, the final information matrix can be written as,
$$
I(\hat{\bm{\gamma}}) = \ddot{Q}(\bm{\gamma}|\hat{\gamma}) - \left\{\Big[\sum_{i = 1}^n \sum_{j = 1}^{J} \hat{w}_{iy_i}S_i(\hat{\bm{\gamma}})S_i(\hat{\bm{\gamma}}^T)\Big] - \sum_{i = 1}^n \dot{q}_i(\bm{\gamma}|\hat{\bm{\gamma}})\dot{q}_i(\bm{\gamma}|\hat{\bm{\gamma}})^T\right\},
$$
where $ \hat{\bm{\gamma}} $ and $ \hat{w}_{iy_i} $ are the corresponding estimates of  $ \bm{\gamma} $ and $ w_{iy_i} $ respectively after convergence 
and $ S_i(\hat{\bm{\gamma}}) = \left[\frac{\partial l(\bm{\gamma}|\bm{x}_i, y_i, r_i)}{\partial \bm{\gamma}}\right]_{\bm{\gamma} = \bm{\hat{\gamma}}} $. The estimated covariance matrix of $ \hat{\bm{\beta}} $ 
is the upper $ (J-1+p) \times (J-1+p) $ block of $ [I(\hat{\bm{\gamma}})]^{-1} $. Hence, the standard errors $ \hat{s}_{\beta_j} $ of the individual parameter $ \beta_j $ are the square root of the diagonal elements of this covariance matrix 
Similarly, the $ 100(1 -\alpha)\% $ confidence interval can be constructed by $ (\hat{\beta}_j - z_{\alpha/2}\hat{s}_{\beta_j}, \hat{\beta}_j + z_{\alpha/2}\hat{s}_{\beta_j}) $, where $ z_\alpha $ is the upper $ 100(1 - \alpha)\% $ quantile of the standard Normal distribution.

\subsection{Theoretical Convergence of the EM Algorithm} \label{section_convergence_theory}

This section provides an overview of the convergence properties of the proposed EM algorithm. While precise mathematical proofs for convergence are challenging due to the complexity of the ordinal likelihood function, we outline a general framework to support the validity of the inferences derived from our approach.

Wu \cite{wu1983convergence} and Boyles \cite{boyles1983convergence} established general conditions for the convergence of the EM algorithm, including the unimodality of the likelihood and other regularity assumptions. Vaida \cite{vaida2005parameter} noted that verifying these conditions can be impractical in applied settings and proposed an alternative, simpler condition: demonstrating that $Q(\cdot|\cdot)$ has a unique global maximum. In addition, Vaida emphasized that the utility of the EM algorithm depends on regularity conditions typically satisfied by exponential families, including cumulative logistic (proportional odds) regression models. Without such conditions, the EM algorithm would have limited practical utility.

To establish that $Q(\cdot|\cdot)$ has a unique global maximum, we draw upon the theory of surrogate loss functions. Bartlett et al. \cite{bartlett2006convexity} and Agarwal \cite{agarwal2008generalization} introduced surrogate functions for minimizing the 0-1 loss and proved that these functions are Fisher consistent. Pedregosa et al. \cite{pedregosa2017consistency} extended this work to ordinal regression models, showing that the surrogate loss functions are Fisher consistent for a broad class of models, including cumulative logit models. Fisher consistency, broadly speaking, ensures that minimizing the surrogate loss function corresponds to achieving the Bayes-optimal risk. While rigorous definitions of Fisher consistency are beyond the scope of this discussion, Pedregosa et al. \cite{pedregosa2017consistency} provide comprehensive details.

Theorem 7 in Pedregosa et al. \cite{pedregosa2017consistency} states that the cumulative logit surrogate loss function is Fisher consistent. This result is critical, as it provides an indirect justification for the existence of a global maximum for $Q(\cdot|\cdot)$, given that minimizing a surrogate loss function is equivalent to maximizing a likelihood function. Moreover, the maximization of $Q(\cdot|\cdot)$ involves two separate steps (see Section \ref{section_M_step}): 
(1) Maximizing the cumulative logit likelihood, which trivially satisfies the necessary criteria.
(2) Maximizing the logistic regression likelihood, which also satisfies these criteria as the logistic model is a special case of the cumulative logistic model with two categories.

\section{Numerical Studies}  \label{section_simulation}

Section \ref{sec:intro} introduced a motivating example from a Phase III clinical trial. Considering the sample sizes typical of different phases of clinical trials and the corresponding models for hypothesis testing, this section presents simulation studies designed to mimic these scenarios. The simulations evaluate the operating characteristics of the proposed EM algorithm under varying sample sizes, missingness levels, and data structures. The models and simulation settings are summarized in Table \ref{table_simulation_setting}.

\subsection{Simulation Settings}
The variable $X_1$ represents the treatment indicator, randomized between the active and control groups with a randomization ratio of 2:1 (67\% active and 33\% control). Additional covariates $X_2$, $X_3$, and $X_4$ are generated to represent common predictors in clinical trials, ensuring realistic data structures for simulations. These variables and the response variable $Y$ are summarized in Table \ref{table_simulation_setting}.

We consider five scenarios with sample sizes \( n = 60, 150, 250, 500, 1000 \). In each scenario, 1000 simulation runs are performed. The covariates \( X_1, X_2, X_3, X_4 \) are generated as follows:
\[
X_1 \sim \text{Bernoulli}(0.67), \quad X_2 \sim \text{Bernoulli}(0.30), \quad X_3 \sim \text{Gamma}(17, 0.2), \quad X_4 \sim \text{Lognormal}(3.1, 0.65).
\]

The ordinal response variable \( Y \) is generated using the cumulative logit model:
\bea  \label{equation_PO_regression_simulation}
\log\left(\frac{P(Y_i \leq j)}{1 - P(Y_i \leq j)}\right) &=& \beta_{0j} + \beta_1 X_1 + \beta_2 X_3 + \beta_3 X_4,
\label{eq:simlset1}
\eea
where \( Y \in \{1, 2, 3\} \). The true parameter values for \( \bm{\beta} = (\beta_{01}, \beta_{02}, \beta_1, \beta_2, \beta_3) \) are set to \( (1, -0.6, -1, 0.005, -0.100)^T \). Additional simulations with \( Y \) taking 5 categories are provided in the Supplementary Material.

Note that, using equation (\ref{eq:simlset1}), the cumulative probabilities \( P(Y_i \leq j) \) for \( j = 1, 2, \ldots, J \) are calculated. The category-specific probabilities \( \pi_{ij} = P(Y_i = j | \ux_i) \) are then obtained as:
\[
\pi_{ij} = P(Y_i \leq j) - P(Y_i \leq j-1),
\]
where \( P(Y_i \leq 0) = 0 \) and \( P(Y_i \leq J) = 1 \). Finally, the ordered multinomial responses \( Y_i \) are generated by drawing from a multinomial distribution with probabilities \( \pi_{ij} \) for each category \( j \).

Missing data in \( Y \) are induced by generating the missingness indicator \( R \) using:
\[
\log\left(\frac{P(R_i = 1)}{P(R_i = 0)}\right) = \alpha_0 + \alpha_1 X_1 + \alpha_2 X_2 + \alpha_3 X_3 + \alpha_4 X_4 + \alpha_5 Y_i,
\]
where the proportion of missing data is determined by the parameter vector \( \ualpha \). We examine three scenarios with approximately 10\%, 25\%, and 45\% missing data. 

Four estimation methods are evaluated:
\begin{itemize}
    \item \textbf{Whole}: Proportional odds regression on the complete dataset without missing values.
    \item \textbf{CC}: Complete case analysis, where missing values are excluded.
    \item \textbf{EM}: The proposed EM algorithm as described in Section \ref{section_method}.
    \item \textbf{MI}: Multiple imputation using the \textsf{R} package \package{mice} \citep{van2011mice}, assuming the missing data mechanism is MAR.
\end{itemize}
\textcolor{black} {The multiple imputation approach implemented in \package{mice} adopts multivariate imputation by chained equations (MICE), which iteratively updates missing values by sampling from their conditional distributions. This process effectively leverages Gibbs sampling to quickly achieve stationarity, typically within a small number of iterations. In this study, the \package{mice} package was used with its default settings: five imputations and five iterations. The imputation process employs predictive mean matching, and pooled parameter estimates are obtained by fitting proportional odds models to the imputed datasets.}

\subsection{Operating Characteristics in Simulation Studies}

\begin{figure}[ht]
	\centering
	\includegraphics[width = \textwidth, height = 10 cm]{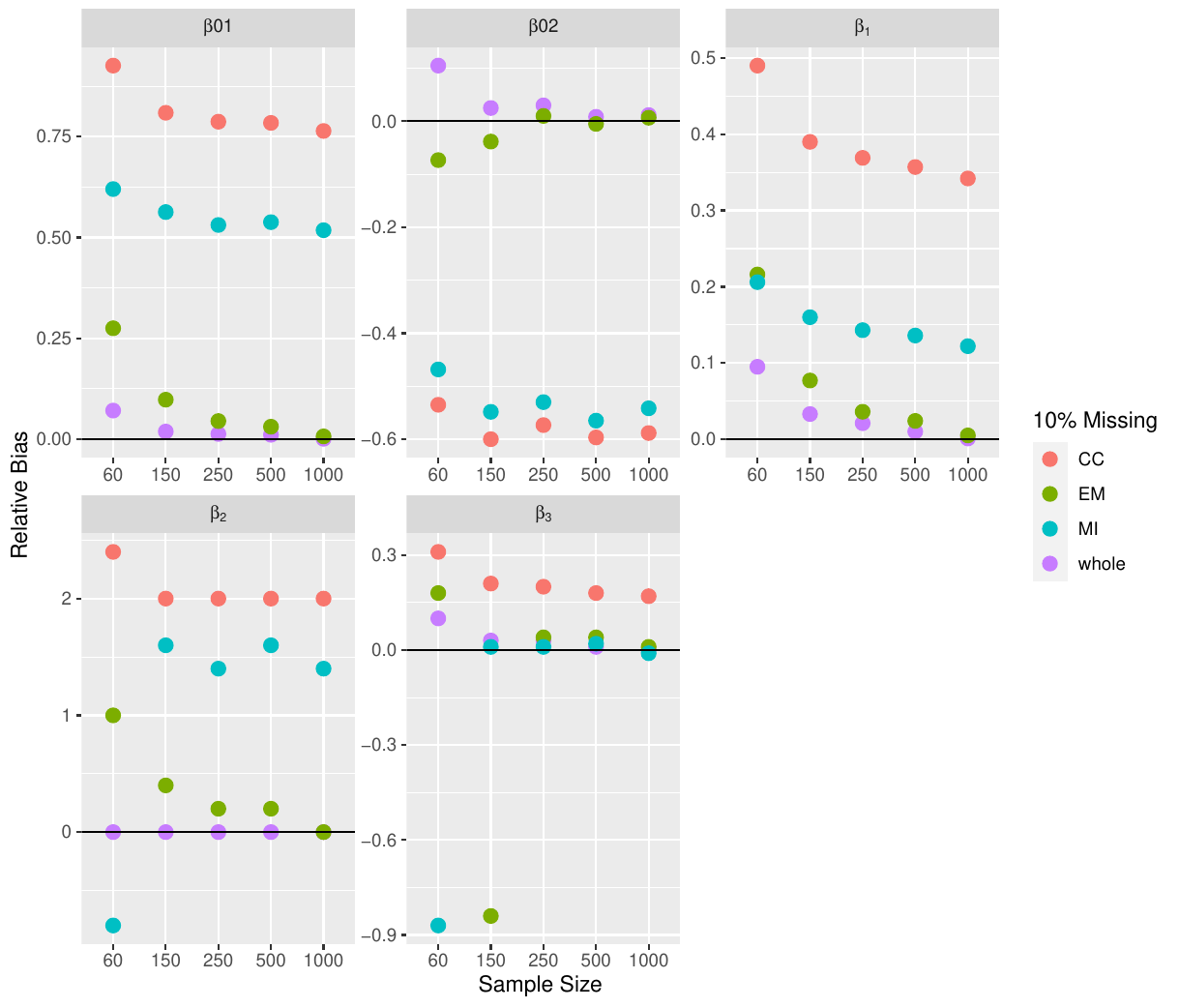}
	\caption{The relative biases of mean estimates of $ \beta $ from 1000 replications using different methods along with the true values when about 10\% data are missing. A horizontal line with zero relative bias has been added for reference. \label{figure_simulation_missing_10}}
\end{figure}

\begin{figure}[ht]
	\centering
	\includegraphics[width = \textwidth, height = 10 cm]{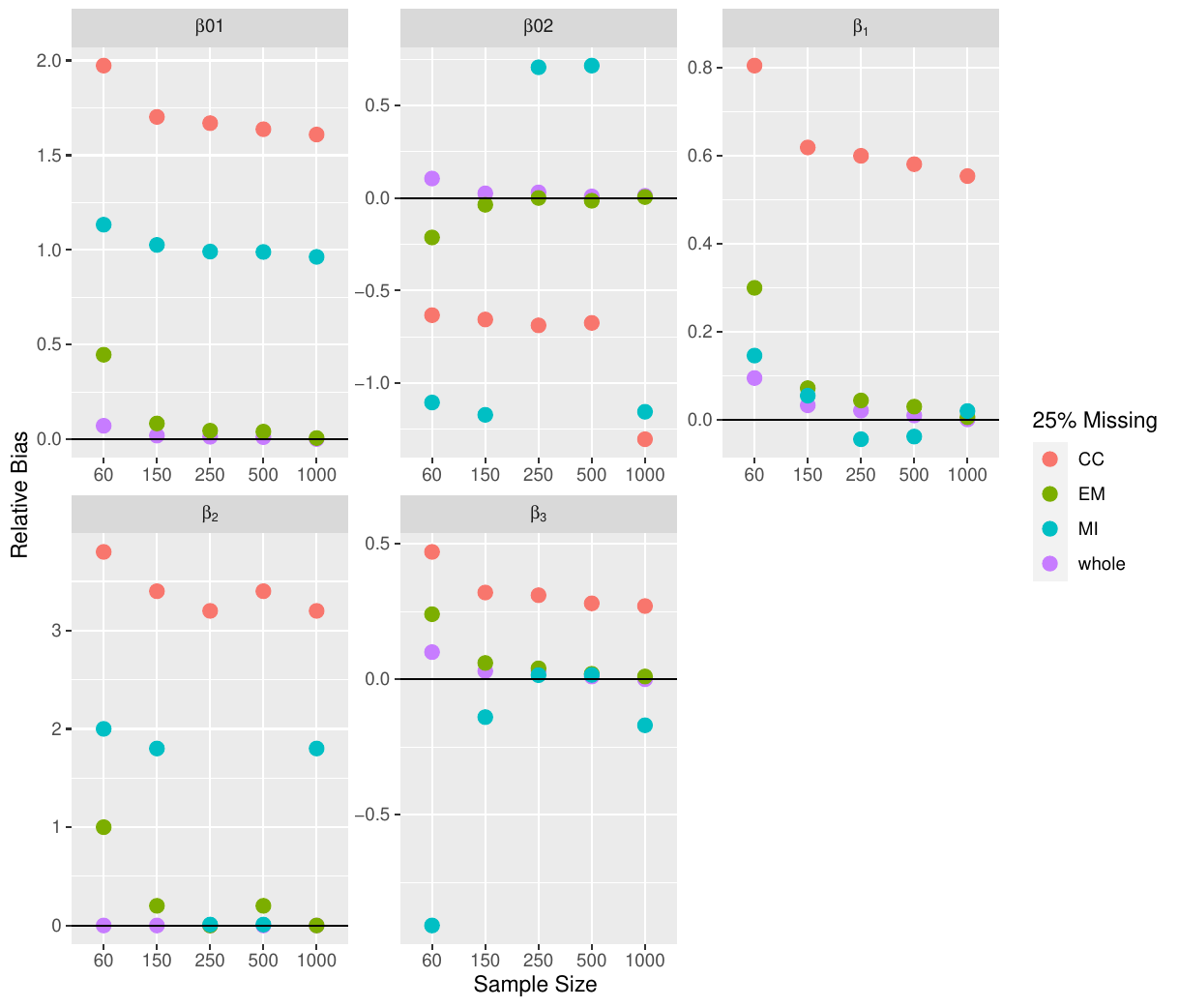}
	\caption{The relative biases of mean estimates of $ \beta $ from 1000 replications using different methods along with the true values when about 25\% data are missing. A horizontal line with zero relative bias has been added for reference. \label{figure_simulation_missing_25}}
\end{figure}

\begin{figure}[ht]
	\centering
	\includegraphics[width = \textwidth, height = 10 cm]{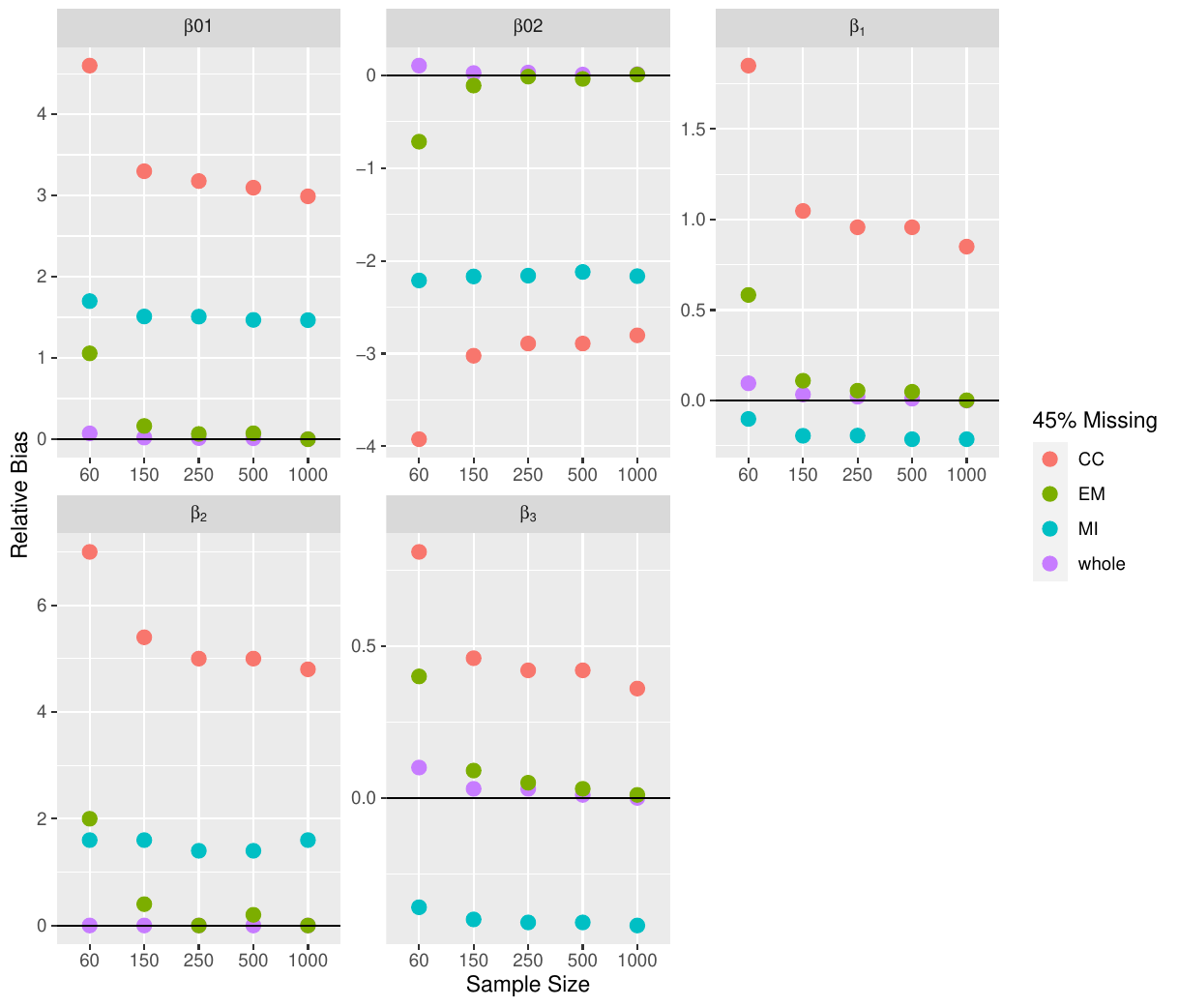}
	\caption{The relative biases of mean estimates of $ \beta $ from 1000 replications using different methods along with the true values when about 45\% data are missing. A horizontal line with zero relative bias has been added for reference. \label{figure_simulation_missing_45}}
\end{figure}

Our primary interest lies in estimating \( \ubeta \), particularly the coefficient of the treatment variable \( X_1 \), as it enables a direct comparison of treatment effects. Tables \ref{table_sim1mis1001}--\ref{table_sim3mis45} and Figures \ref{figure_simulation_missing_10}--\ref{figure_simulation_missing_45} summarize key operating characteristics for the various methods across different sample sizes and levels of missingness, after excluding non-converged runs. The following summary statistics are reported:

\begin{itemize}
    \item \( E[\hat{\beta}] \): Mean of the 1000 parameter estimates; smaller differences between this value and the true \( \beta \) are desirable.
    \item Absolute bias: \( |E[\hat{\beta}] - \beta| \); smaller values indicate better performance.
    \item MSE: Mean squared error, defined as \(\text{bias}^2 + \text{SD}^2\); smaller values are preferable.
    \item 95\% CP: Coverage probability of the 95\% confidence interval; closer to 0.95 is ideal.
    \item Relative bias: Defined as \( (E[\hat{\beta}] - \beta)/\beta \); values close to 0 indicate minimal bias.
\end{itemize}

\textcolor{black}{Simulation results for the different methods are presented as follows: Table \ref{table_sim1mis1001} and Figure \ref{figure_simulation_missing_10} summarize results when approximately 10\% of responses are missing, Table \ref{table_sim3mis2501} and Figure \ref{figure_simulation_missing_25} present results for 25\% missing data, and Table \ref{table_sim3mis45} and Figure \ref{figure_simulation_missing_45} display outcomes for 45\% missing responses.}

As expected, the "Whole" dataset approach (i.e., no missing data) performs best across all metrics. Among the methods for handling missing data, the proposed "EM" algorithm consistently outperforms "CC" (complete case analysis) and "MI" (multiple imputation) in terms of reducing bias and MSE. For instance, when \( n = 150 \) and approximately 10\% of the data are missing, the true value of \( \beta \) for the treatment covariate \( X_1 \) is \(-1\). The estimated values for "CC," "EM," and "MI" are \(-1.390\), \(-1.077\), and \(-1.160\), respectively. The corresponding absolute biases are 0.390, 0.077, and 0.160, while the MSE values are 0.3557, 0.1869, and 0.1911 (see Table \ref{table_sim1mis1001}). These results demonstrate the superiority of the proposed "EM" algorithm in handling nonignorable missingness, compared to methods assuming MAR or excluding missing data entirely.

Similar trends are observed for higher levels of missingness. For instance, when 25\% or 45\% of responses are missing (Tables \ref{table_sim3mis2501} and \ref{table_sim3mis45}, respectively), the "EM" method continues to yield lower bias and MSE compared to "CC" and "MI." These findings highlight the robustness of the proposed algorithm in addressing nonignorable missing data.

The 95\% CP of the "EM" method approaches 0.95 as the sample size \( n \) increases, providing evidence of the consistency and convergence properties of the estimator. Furthermore, the bias of the "EM" estimates decreases with increasing sample size, as demonstrated in Tables \ref{table_sim1mis1001}, \ref{table_sim3mis2501}, and \ref{table_sim3mis45}. This observation further supports the theoretical consistency of the proposed method.

Figures \ref{figure_simulation_missing_10}--\ref{figure_simulation_missing_45} illustrate the relative bias of the estimated parameters for different methods compared to the true parameter values. The "EM" estimates exhibit relative biases closer to 0 than "CC" and "MI," further underscoring the effectiveness of the proposed approach in maintaining accuracy under varying levels of missingness.

\textcolor{black}{While MICE is widely used in practice due to its flexibility and ease of implementation, it is essential to base the choice of imputation method on underlying assumptions and data characteristics. In scenarios with nonignorable missing data, MICE may not adequately account for the missingness mechanism, as it inherently assumes the MAR framework. This highlights the importance of adopting methods, such as the proposed EM algorithm, that directly address MNAR settings.}

\newpage 

\section{Example: Revisit of Plaque Psoriasis Study}  \label{section_real_data}
This section revisits the Phase III psoriasis study dataset introduced in Section \ref{sec:intro}, which evaluates the efficacy of a treatment for moderate-to-severe plaque psoriasis. The Physician’s Global Assessment (PGA), measured on an ordinal five-point scale ranging from $0$ to $4$, serves as the primary endpoint. As previously noted, 13.5\% of the PGA responses were missing. In prior analyses, the PGA scores were dichotomized as  $\text{PGA} \leq 1$  versus  $\text{PGA} > 1$ , and a binary logistic regression model was fitted using a complete case (CC) approach. However, this method excluded missing responses, leading to potential bias and information loss.\\

In this analysis, we consider the fully observed ordinal PGA scores as the response variable  Y  instead of dichotomizing them. To better capture the ordinal nature of the data and minimize information loss, we fit the following cumulative logit model to determine the clinical relevance of predictors in explaining PGA scores:

\be
\log\left(\frac{P(Y_i \le j)}{1- P(Y_i \le j)}\right) = \beta_{0j} + \beta_1 \text{Treatment}_i + \beta_2 \text{Age (in year)}_i + \beta_3 \text{Sex}_i + \beta_4 \text{Weight}_i + \beta_5 \text{onsetage}_i,
\label{eq:motivatingex}
\ee
j = 0, 1, 2, 3.

In order to fit the above model with the proposed method, we fit the corresponding \textcolor {black} {logit($P(R = 1)$)} model as the following:
\be \label{equation_missing_trial}
\log\left(\frac{P(R_i = 1)}{P(R_i = 0)}\right) = \alpha_0 + \alpha_1 \text{Treatment}_i + \alpha_2 \text{Age (in year)}_i + \alpha_3 \text{Sex}_i + \alpha_4 \text{Weight}_i + \alpha_5 \text{onsetage}_i + \alpha_6 Y_i
\ee

In the above model (\ref{eq:motivatingex}, the variables Treatment (including three treatment levels -- doses 5MG, 10MG and Placebo), Age (in years), Sex (Male or Female), Weight and Onsetage (number of years since the fist disease diagnosis) were considered because of their clinical relevance. After fitting the proposed model, the results corresponding the model of equation (\ref{equation_missing_trial}) are shown in Table~\ref{table_logit_r_Y_EM}.

\textcolor {black} {Notice that in Table \ref{table_logit_r_Y_EM}, the estimates corresponding to the variable $Y$ is significant (p-value 0.021) at 5\% level of condfidence with the 95\% confidence interval (-2.281, -0.184), implying the missingness of $Y$ may be nonignorable.} 
The Table \ref{tab:table_tofareslt} shows odds ratio estimates fitting a CC-analysis model, proposed method using EM-algorithm, and the same model using the MI approach. Notice that the estimates of the treatments 5 MG and 10 MG are highly significant (corresponding p-values are 0.000 and 0.000, respectively) compared to the placebo arm using all three approaches, which is consistent with the study results \citep{papp2015tofacitinib}. One more interesting fact that the p-value (= 0.034) corresponding to the variable ONSETAGE is statistically significant at the 5\% level, when the same is not significant (p-value = 0.059) from the complete case analysis removing all missing observations or using the MI approach (p-value = 0.253). The variable ONSETAGE (number of years since first diagnosed) is a clinically important factor, as patients with early diagnosed of psoriasis may respond to the treatment differently than the patients with late diagnosed \citep{theodorakopoulou2016early, singh2018effect}.

\section{Discussion and Future Considerations} \label{section_discussion}

This study introduced a new EM algorithm for cumulative logit ordinal regression under nonignorable missing responses, demonstrating its theoretical validity and practical effectiveness. The algorithm leverages the proportional odds framework to provide a robust solution for ordinal data analysis in the presence of nonignorable missingness mechanisms.

The convergence properties of the EM algorithm, detailed in Section \ref{section_convergence_theory}, are supported by theoretical guarantees such as the Fisher consistency of the cumulative logit surrogate loss function \citep{pedregosa2017consistency}. These results ensure that the algorithm converges to stationary points of the joint likelihood, provided the necessary regularity conditions are met. Simulation studies further confirmed these theoretical findings, showing consistent convergence and superior performance of the EM algorithm compared to traditional methods such as complete case (CC) analysis and multiple imputation (MI). The proposed method achieved lower bias and mean squared error (MSE) across a range of missingness mechanisms and proportions, particularly excelling under MNAR scenarios.

Despite its strengths, the application of the EM algorithm introduces practical challenges, particularly in model selection. \textcolor {black} {In the simulation studies, the missing response model included additional covariates beyond those used to generate the response variable. This reflects realistic scenarios where missingness mechanisms may depend on unobserved or auxiliary information. Moreover, it underscores the importance of rigorous variable selection procedures to ensure reliable inferences.} Additionally, the complexity of the response structure plays a critical role; models with fewer response categories demonstrated better performance due to reduced parameter estimation complexity. For instance, the EM algorithm produced more precise estimates with three-category responses compared to five-category responses, highlighting the need to carefully consider response structure in practice.

The real-world application to the psoriasis dataset further demonstrated the utility of the proposed method. Significant p-values and confidence interval estimates indicated the presence of nonignorable missingness, validating the robustness of the EM algorithm. However, it is important to recognize that statistical significance does not necessarily imply clinical relevance. Collaborating with domain experts is essential to ensure that the results are interpreted in the context of real-world decision-making.

This work also opens several avenues for future research. The proposed method could be extended to accommodate alternative missing data mechanisms, such as pattern-mixture models or hybrid MAR-MNAR frameworks. Additionally, generalizing the algorithm to more complex models, such as generalized ordinal regression or mixed-effects frameworks, would enhance its applicability. Future work could also explore computationally efficient adaptations for large-scale datasets and regularization techniques for handling high-dimensional covariates. Comparing the EM algorithm with sensitivity analyses under MAR assumptions and refining MICE algorithm settings could provide further insights into its relative strengths and limitations.

In conclusion, the proposed EM algorithm offers a theoretically sound and practically robust approach for addressing nonignorable missing responses in cumulative logit models. By integrating theoretical guarantees with strong empirical performance, this study provides a foundation for advancing ordinal data analysis in both research and applied settings.

\section*{Data Availability}

The data that support the findings of this study are available from Pfizer. Restrictions apply to the availability of these data, which were used under license for this study. Data are available from the authors with the permission of Pfizer.

\section*{Acknowledgements} The clinical trial reported in this article has been sponsored by Pfizer.  We are indebted to the editor, the associate editor, and the referees for their constructive comments which helped to improve the article considerably.


\bibliography{referencePOmissing.bib}


\newpage

\begin{table}[ht]
{\scriptsize
	\centering
	\caption{Simulation Settings. \label{table_simulation_setting}}
	\begin{tabular}{c|l|l}
		\hline
		\hline
		Variable & Description and Model & Parameter  \\
		\hline
		\hline
		$ \bm{X} $ & $ X_1: $ Treatment (fixed); (1/3 are 0 \& 2/3 are 1) & \\
        & $ X_2: $ Bernoulli (0.3) &  \\
		& $ X_3: $ Gamma(shape = 17, rate = 0.2) &  \\
		& $ X_4: $ Lognormal(mean = 3.1, sd = 0.65) &  \\
		\hline
		Y & $ J = 3 $ and $ Y \in \{1, 2, 3\} $ & $ \bm{\beta}_{\text{true}} = (1, -0.6, -1, 0.005, -0.1)^T $  \\
		\hline
		R & $ Y $ is included as a predictor & Simulation 1: $ \alpha_{\text{true}} = (1, -2, -0.6, 0.05, -0.1, -4)^T $  \\
		& & (about 10\% data are missing).  \\
		& & Simulation 2: $ \alpha_{\text{true}} = (2.8, -2, -0.6, 0.05, -0.1, -4)^T $  \\
		& & (about 25\% data are missing).  \\
		& & Simulation 3: $ \alpha_{\text{true}} = (4.8, -2, -0.6, 0.05, -0.1, -4)^T $  \\
		& & (about 45\% data are missing).  \\
		\hline
		\hline
	\end{tabular}
}
\end{table}

    
\begin{table}[ht]
	{\tiny
		\centering
		\caption{Simulation 1: with $\sim$ 10\% missing responses: Simulation results for $n=60, 150,250,500$ with $1000$ replications for each scenario. Summary statistics, (1) $E[\widehat{\beta}]$ is the mean of $1000$ estimators, (2) Absolute bias = $| E[\hat{\beta}]-\beta |$ (3) MSE = bias$^2$+SD$^2$, (4) 95\% CP is the 95\% coverage percentage. The best values among ``CC", ``EM", and ``"MI" are marked in bold.  \label{table_sim1mis1001}}
		\begin{tabular}{rrrrrrrrrrrrrrrrrrr}
			\hline
			&       & {\bf } & \multicolumn{4}{|c}{{\bf $E[\widehat{\beta}]$}} & \multicolumn{ 4}{|c}{{\bf Absolute Bias}} & \multicolumn{ 4}{|c}{{\bf MSE}} & \multicolumn{ 4}{|c}{{\bf 95\%CP}}  \\
			&       & {\bf \ubeta} & {\bf whole} & {\bf CC} & {\bf EM} & {\bf MI} & {\bf whole} & {\bf CC} & {\bf EM} & {\bf MI} & {\bf whole} & {\bf CC} & {\bf EM} & {\bf MI} & {\bf whole} & {\bf CC} & {\bf EM} & {\bf MI} \\
			\hline
			\multicolumn{ 1}{c}{$n=60$} & $y\le 1$  & 1.000 & 1.071 & 1.926 & 1.275 & 1.620 & 0.071 & 0.926 & \textbf{0.275} & 0.620 & 0.3552 & 1.3276 & \textbf{0.6894} & 0.8317 & 0.952 & 0.792 & \textbf{0.881} & 0.824 \\
			\multicolumn{ 1}{c}{} & $y\le 2$  & -0.600 & -0.663 & -0.279 & -0.556 & -0.319 & 0.063 & 0.321 & \textbf{0.044} & 0.281 & 0.3302 & 0.5397 & 0.4808 & \textbf{0.4614} & 0.951 & \textbf{0.927} & 0.924 & 0.891 \\
			\multicolumn{ 1}{c}{} & $x_{1}= 1$  & -1.000 & -1.095 & -1.490 & -1.216 & -1.206 & 0.095 & 0.490 & 0.216 & \textbf{0.206} & 0.4590 & 0.8709 & \textbf{0.6465} & 0.5093 & 0.945 & 0.931 & 0.931 & \textbf{0.940} \\
			\multicolumn{ 1}{c}{} & $x_3$    & 0.005 & 0.005 & 0.017 & 0.010 & 0.001 & 0.000 & 0.012 & 0.008 & \textbf{0.005} & 0.0002 & 0.0004 & \textbf{0.0003} & \textbf{0.0003} & 0.958 & 0.914 & \textbf{0.937} & 0.909 \\
			\multicolumn{ 1}{c}{} & $x_4$    & -0.100 & -0.110 & -0.131 & -0.118 & -0.013 & 0.010 & 0.031 & 0.018 & \textbf{0.008} & 0.0010 & 0.0023 & 0.0016 & \textbf{0.0009} & 0.951 & 0.914 & \textbf{0.948} & 0.933  \\
			&       &       &       &       &       &       &       &       &       &       &      &      &    \\
			\multicolumn{ 1}{c}{$n=150$} & $y\le 1$  & 1.000 & 1.019 & 1.809 & 1.098 & 1.563 & 0.019 & 0.809 & \textbf{0.098} & 0.563 & 0.1183 & 0.8126 & \textbf{0.1967} & 0.5012 & 0.950 & 0.495 & \textbf{0.934} & 0.623 \\
			\multicolumn{ 1}{c}{} & $y \le 2$  & -0.600 & -0.615 & -0.240 & -0.577 & -0.271 & 0.015 & 0.360 & \textbf{0.023} & 0.329 & 0.1069 & 0.2682 & \textbf{0.1254} & 0.2637 & 0.954 & 0.838 & \textbf{0.951} & 0.778\\
			\multicolumn{ 1}{c}{} & $x_{1}= 1$  & -1.000 & -1.033 & -1.390 & -1.077 & -1.160 & 0.033 & 0.390 & \textbf{0.077} & 0.160 & 0.1601 & 0.3557 & \textbf{0.1869} & 0.1911 & 0.949 & 0.874 & \textbf{0.942} & 0.927  \\
			\multicolumn{ 1}{c}{} & $x_3$    & 0.005 & 0.005 & 0.015 & 0.007 & 0.013 & 0.000 & 0.010 & \textbf{0.002} & 0.008 & 0.0001 & 0.0002 & \textbf{0.0001} & 0.0002 & 0.949 & 0.830 & \textbf{0.940} & 0.840 \\
			\multicolumn{ 1}{c}{} & $x_4$    & -0.100 & -0.103 & -0.121 & -0.106 & -0.101 & 0.003 & 0.021 & 0.006 & \textbf{0.001} & 0.0003 & 0.0008 & \textbf{0.0004} & \textbf{0.0004} & 0.956 & 0.872 & 0.959 & \textbf{0.950} \\
			&       &       &       &       &       &       &       &       &       &       &  \\
			\multicolumn{ 1}{c}{$n=250$} & $y\le 1$  & 1.000 & 1.013 & 1.787 & 1.045 & 1.531 & 0.013 & 0.787 & \textbf{0.045} & 0.531 & 0.0669 & 0.7069 & \textbf{0.0932} & 0.4131 & 0.957 & 0.265 & \textbf{0.946} & 0.489 \\
			\multicolumn{ 1}{c}{} & $y\le 2$  & -0.600 & -0.618 & -0.256 & -0.606 & -0,282 & 0.018 & 0.344 & \textbf{0.006} & 0.318 & 0.0630 & 0.1985 & \textbf{0.0703} & 0.2079 & 0.954 & 0.763 & \textbf{0.947} & 0.708 \\
			\multicolumn{ 1}{c}{} & $x_{1}=1$  & -1.000 & -1.021 & -1.369 & -1.036 & -1.143 & 0.021 & 0.369 & \textbf{0.036} & 0.143 & 0.0844 & 0.2450 & \textbf{0.0938} & 0.1132 & 0.954 & 0.824 & 0.953 & 0.925 \\
			\multicolumn{ 1}{c}{} & $x_3$    & 0.005 & 0.005 & 0.015 & 0.006 & 0.012 & 0.000 & 0.010 & \textbf{0.001} & 0.007 & 0.0000 & 0.0002 & \textbf{0.0000} & 0.0001 & 0.944 & 0.739 & \textbf{0.956} & 0.770 \\
			\multicolumn{ 1}{c}{} & $x_4$    & -0.100 & -0.103 & -0.120 & -0.104 & -0.101 & 0.003 & 0.020 & 0.004 & \textbf{0.001} & 0.0002 & 0.0006 & \textbf{0.0002} & \textbf{0.0002} & 0.939 & 0.771 & 0.937 & \textbf{0.948} \\
			&       &       &       &       &       &       &       &       &       &       &  \\
			\multicolumn{ 1}{c}{$n=500$} & $y\le 1$  & 1.000 & 1.011 & 1.784 & 1.031 & 1.538 & 0.011 & 0.784 & \textbf{0.031} & 0.538 & 0.0354 & 0.6621 & \textbf{0.0474} & 0.3809 & 0.943 & 0.044 & \textbf{0.945} & 0.290 \\
			\multicolumn{ 1}{c}{} & $y\le 2$  & -0.600 & -0.605 & -0.242 & -0.597 & -0.261 & 0.005 & 0.358 & \textbf{0.003} & 0.339 & 0.0318 & 0.1689 & \textbf{0.0329} & 0.1872 & 0.942 & 0.544 & \textbf{0.945} & 0.495 \\
			\multicolumn{ 1}{c}{} & $x_{1}=1$  & -1.000 & -1.010 & -1.357 & -1.024 & -1.136 & 0.010 & 0.357 & \textbf{0.024} & 0.136 & 0.0437 & 0.1818 & \textbf{0.0492} & 0.0694 & 0.944 & 0.683 & \textbf{0.941} & 0.885 \\
			\multicolumn{ 1}{c}{} & $x_3$    & 0.005 & 0.005 & 0.015 & 0.006 & 0.013 & 0.000 & 0.010 & \textbf{0.001} & 0.008 & 0.0000 & 0.0001 & \textbf{0.0000} & 0.0001 & 0.942 & 0.490 & \textbf{0.930} & 0.615 \\
			\multicolumn{ 1}{c}{} & $x_4$    & -0.100 & -0.101 & -0.118 & -0.099 & 0.102 & 0.001 & 0.018 & \textbf{0.002} & 0.001 & 0.957 & 0.618 & 0.958 & \textbf{0.951} \\
   	&       &       &       &       &       &       &       &       &       &       &  \\
			\multicolumn{ 1}{c}{$n= 1000$} & $y\le 1$  & 1.000 & 1.001 & 1.764 & 1.007 & 1.518 & 0.001 & 0.764 & \textbf{0.007} & 0.518 & 0.0161 & 0.6041 & \textbf{0.0201} & 0.3383 & 0.955 & 0.000 & \textbf{0.956} & 0.158  \\
			\multicolumn{ 1}{c}{} & $y\le 2$  & -0.600 & -0.607 & -0.247 & -0.604 & -0.275 & 0.007 & 0.353 & \textbf{0.004} & 0.325 & 0.0155 & 0.1443 & \textbf{0.0159} & 0.1543 & 0.946 & 0.283 & \textbf{0.948} & 0.344 \\
			\multicolumn{ 1}{c}{} & $x_{1}=1$  & -1.000 & -1.001 & -1.342 & -1.005 & -1.122 & 0.001 & 0.342 & \textbf{0.005} & 0.122 & 0.0199 & 0.1421 & \textbf{0.0218} & 0.0418 & 0.949 & 0.452 & \textbf{0.952} & 0.844 \\
			\multicolumn{ 1}{c}{} & $x_3$    & 0.005 & 0.005 & 0.015 & 0.005 & 0.012 & 0.000 & 0.010 & \textbf{0.000} & 0.007 & 0.0000 & 0.0001 & \textbf{0.0000} & 0.0001 & 0.947 & 0.225 & \textbf{0.955} & 0.428  \\
			\multicolumn{ 1}{c}{} & $x_4$    & -0.100 & -0.100 & -0.117 & -0.101 & -0.099 & 0.000 & 0.017 & \textbf{0.001} & \textbf{0.001} & 0.0000 & 0.0003 & 0.0001 & \textbf{0.0000} & 0.940 & 0.344 & \textbf{0.934} & 0.910  \\
			\hline
		\end{tabular}
		
	}
\end{table}


    
\begin{table}[ht]
\centering
\caption{Simulation 2: with $\sim$ 25\% missing responses: Simulation results for $n=60, 150,250,500$ with $1000$ replications for each scenario. Summary statistics (1) $E[\widehat{\beta}]$ is the mean of $1000$ estimators (2) Absolute bias = $| E[\hat{\beta}]-\beta |$ (3) MSE = bias$^2$+SD$^2$ (4) 95\% CP is the 95\% coverage percentage. The best values among ``CC", ``EM", and ``"MI" are marked in bold. \label{table_sim3mis2501}}
{\tiny
	\begin{tabular}{rrrrrrrrrrrrrrrrrrr}
			\hline
			&       & {\bf } & \multicolumn{4}{|c}{{\bf $E[\widehat{\beta}]$}} & \multicolumn{ 4}{|c}{{\bf Absolute Bias}} & \multicolumn{ 4}{|c}{{\bf MSE}} & \multicolumn{ 4}{|c}{{\bf 95\%CP}} \\
			&       & {\bf \ubeta} & {\bf whole} & {\bf CC} & {\bf EM} & {\bf MI} & {\bf whole} & {\bf CC} & {\bf EM} & {\bf MI} & {\bf whole} & {\bf CC} & {\bf EM} & {\bf MI} & {\bf whole} & {\bf CC} & {\bf EM} & {\bf MI} \\
			\hline
		\multicolumn{ 1}{c}{n=60} & $ y\le 1 $ & 1.000 & 1.071 & 2.973 & 1.446 & 2.133 & 0.071 & 1.973 & \textbf{0.446} & 1.133 & 0.3552 & 5.1162 & \textbf{1.6538} & 2.0336 & 0.952 & 0.451 & \textbf{0.899} & 0.560 \\
		\multicolumn{ 1}{c}{} & $ y\le 2 $ & -0.600 & -0.663 & -0.220 & -0.472 & 0.063 & 0.820 & 1.133 & \textbf{0.128} & 0.739 & 0.3302 & 1.5607 & \textbf{0.8903} & 1.1390 & 0.951 & 0.837 & \textbf{0.928} & 0.658 \\
		\multicolumn{ 1}{c}{} & $ x_1 = 1 $  & -1.000 & -1.095 & -1.805 & -1.300 & -1.146 & 0.095 & 0.805 & 0.300 & \textbf{0.146} & 0.4590 & 1.9163 & 1.1524 & \textbf{0.5145} & 0.945 & 0.912 & 0.942 & \textbf{0.945} \\
		\multicolumn{ 1}{c}{} & $ x_3 $    & 0.005 & 0.005 & 0.024 & 0.010 & 0.015 & 0.000 & 0.019 & \textbf{0.005} & 0.010 & 0.0002 & 0.0009 & \textbf{0.0004} & \textbf{0.0004} & 0.958 & 0.884 & \textbf{0.928} & 0.891 \\
		\multicolumn{ 1}{c}{} & $ x_4 $    & -0.100 & -0.110 & -0.147 & -0.124 & -0.091 & 0.010 & 0.047 & 0.024 & \textbf{0.009} & 0.0010 & 0.0043 & 0.0025 & \textbf{0.0008} & 0.951 & 0.895 & \textbf{0.934} & 0.877 \\
		&       &       &       &       &       &       &       &       &       &       &  \\
		\multicolumn{ 1}{c}{n=150} & $ y\le 1 $ & 1.000 & 1.019 & 2.702 & 1.083 & 2.026 & 0.019 & 1.702 & \textbf{0.083} & 1.026 & 0.1183 & 3.1716 & \textbf{0.2753} & 1.4442 & 0.950 & 0.060 & \textbf{0.943} & 0.288 \\
		\multicolumn{ 1}{c}{} & $ y\le 2 $  & -0.600 & -0.615 & -0.206 & -0.578 & 0.103 & 0.015 & 0.806 & \textbf{0.022} & 0.703 & 0.1069 & 0.8648 & \textbf{0.1376} & 0.7549 & 0.954 & 0.539 & \textbf{0.955} & 0.433 \\
		\multicolumn{ 1}{c}{} & $ x_1 = 1 $  & -1.000 & -1.033 & -1.619 & -1.072 & -1.055 & 0.033 & 0.619 & 0.072 & \textbf{0.055} & 0.1601 & 0.6831 & 0.2257 & \textbf{0.1811} & 0.949 & 0.815 & \textbf{0.946} & 0.931 \\
		\multicolumn{ 1}{c}{} & $ x_3 $  & 0.005 & 0.005 & 0.022 & 0.006 & 0.014 & 0.000 & 0.017 & \textbf{0.001} & 0.009 & 0.0001 & 0.0004 & \textbf{0.0001} & 0.0002 & 0.949 & 0.692 & \textbf{0.950} & 0.804 \\
		\multicolumn{ 1}{c}{} & $ x_4 $   & -0.100 & -0.103 & -0.132 & -0.106 & -0.086 & 0.003 & 0.032 & \textbf{0.006} & 0.014 & 0.0003 & 0.0015 & 0.0005 & \textbf{0.0004} & 0.956 & 0.766 & \textbf{0.964} & 0.797 \\
		&       &       &       &       &       &       &       &       &       &       &  \\
		\multicolumn{ 1}{c}{n=250} & $ y \le 1 $ & 1.000 & 1.013 & 2.669 & 1.045 & 1.991 & 0.013 & 1.669 & \textbf{0.045} & 0.991 & 0.0669 & 2.9314 & \textbf{0.1332} & 1.2898 & 0.957 & 0.003 & \textbf{0.952} & 0.212 \\
		\multicolumn{ 1}{c}{} & $ y \le 2 $  & -0.600 & -0.618 & 0.187 & -0.600 & 0.106 & 0.018 & 0.787 & \textbf{0.000} & 0.706 & 0.0630 & 0.7346 & \textbf{0.0731} & 0.6849 & 0.954 & 0.335 & \textbf{0.959} & 0.276 \\
		\multicolumn{ 1}{c}{} & $ x_1 = 1 $ & -1.000 & -1.021 & -1.600 & -1.044 & -1.044 & 0.021 & 0.600 & \textbf{0.044} & \textbf{0.044} & 0.0844 & 0.5102 & 0.1120 & \textbf{0.1014} & 0.954 & 0.711 & 0.956 & \textbf{0.945} \\
		\multicolumn{ 1}{c}{} & $ x_3 $  & 0.005 & 0.005 & 0.021 & 0.005 & 0.014 & 0.000 & 0.016 & \textbf{0.000} & 0.009 & 0.0000 & 0.0003 & \textbf{0.0001} & 0.0002 & 0.944 & 0.561 & \textbf{0.960} & 0.699 \\
		\multicolumn{ 1}{c}{} & $ x_4 $  & -0.100 & -0.103 & -0.131 & -0.104 & -0.085 & 0.003 & 0.031 & \textbf{0.004} & 0.015 & 0.0002 & 0.0012 & \textbf{0.0003} & 0.0004 & 0.939 & 0.577 & \textbf{0.939} & 0.667 \\
		&       &       &       &       &       &       &       &       &       &       &  \\
		\multicolumn{ 1}{c}{n=500} & $ y\le 1 $  & 1.000 & 1.011 & 2.637 & 1.040 & 1.989 & 0.011 & 1.637 & \textbf{0.040} & 0.989 & 0.0354 & 2.7559 & \textbf{0.0707} & 1.2586 & 0.943 & 0.000 & \textbf{0.943} & 0.106 \\
		\multicolumn{ 1}{c}{} & $ y \le 2 $ & -0.600 & -0.605 & 0.195 & -0.591 & 0.115 & 0.005 & 0.795 & \textbf{0.009} & 0.715 & 0.0318 & 0.6892 & \textbf{0.0379} & 0.6910 & 0.942 & 0.064 & \textbf{0.946} & 0.197 \\
		\multicolumn{ 1}{c}{} & $ x_1 = 1 $  & -1.000 & -1.010 & -1.581 & -1.030 & -1.038 & 0.010 & 0.581 & \textbf{0.030} & 0.038 & 0.0437 & 0.4149 & 0.0590 & 0.0622 & 0.944 & 0.463 & \textbf{0.940} & 0.902 \\
		\multicolumn{ 1}{c}{} & $ x_3 $  & 0.005 & 0.005 & 0.022 & 0.006 & 0.014 & 0.000 & 0.017 & \textbf{0.001} & 0.009 & 0.0000 & 0.0003 & \textbf{0.0000} & 0.0001 & 0.942 & 0.215 & \textbf{0.935} & 0.503 \\
		\multicolumn{ 1}{c}{} & $ x_4 $    & -0.100 & -0.101 & -0.128 & -0.102 & -0.084 & 0.001 & 0.028 & \textbf{0.002} & 0.016 & 0.0001 & 0.0009 & \textbf{0.0001} & 0.0003 & 0.957 & 0.316 & \textbf{0.959} & 0.445 \\
  &       &       &       &       &       &       &       &       &       &       &  \\
		\multicolumn{ 1}{c}{n= 1000} & $ y\le 1 $  & 1.000 & 1.001 & 2.609 & 1.006 & 1.963 & 0.001 & 1.609 & \textbf{0.006} & 0.963 & 0.0161 & 2.6223 & \textbf{0.0297} & 1.1864 & 0.955 & 0.000 & \textbf{0.944} & 0.072 \\
		\multicolumn{ 1}{c}{} & $ y \le 2 $ & -0.600 & -0.607 & 0.182 & -0.603 & 0.093 & 0.007 & 0.782 & \textbf{0.003} & 0.693 & 0.0155 & 0.6367 & \textbf{0.0172} & 0.6237 & 0.946 & 0.000 & \textbf{0.949} & 0.109 \\
		\multicolumn{ 1}{c}{} & $ x_1 = 1 $  & -1.000 & -1.001 & -1.554 & -1.006 & -1.020 & 0.001 & 0.554 & \textbf{0.006} & 0.020 & 0.0199 & 0.3411 & \textbf{0.0253} & 0.0359 & 0.949 & 0.163 & \textbf{0.951} & 0.879 \\
		\multicolumn{ 1}{c}{} & $ x_3 $  & 0.005 & 0.005 & 0.021 & 0.005 & 0.014 & 0.000 & 0.016 & \textbf{0.000} & 0.009 & 0.0000 & 0.0003 & \textbf{0.0000} & 0.0001 & 0.947 & 0.041 & \textbf{0.946} & 0.368 \\
		\multicolumn{ 1}{c}{} & $ x_4 $    & -0.100 & -0.100 & -0.127 & -0.101 & -0.083 & 0.000 & 0.027 & \textbf{0.001} & 0.017 & 0.0000 & 0.0008 & \textbf{0.0001} & 0.0003 & 0.940 & 0.074 & \textbf{0.935} & 0.202 \\
		\hline
	\end{tabular}
}
\end{table}


    
\begin{table}[ht]
\centering
\caption{Simulation 3: with $\sim$ 45\% missing responses: Simulation results for $ n = 60, 150, 250, 500$ with $1000$ replications for each scenario. Summary statistics (1) $E[\widehat{\beta}]$ is the mean of $1000$ estimators (2) Absolute bias = $| E[\hat{\beta}]-\beta |$ (3) MSE = bias$^2$+SD$^2$ (4) 95\% CP is the 95\% coverage percentage. The best values among ``CC", ``EM", and ``"MI" are marked in bold. \label{table_sim3mis45}}
{\tiny
	\begin{tabular}{rrrrrrrrrrrrrrrrrrr}
			\hline
			&       & {\bf } & \multicolumn{4}{|c}{{\bf $E[\widehat{\beta}]$}} & \multicolumn{ 4}{|c}{{\bf Absolute Bias}} & \multicolumn{ 4}{|c}{{\bf MSE}} & \multicolumn{ 4}{|c}{{\bf 95\%CP}} \\
			&       & {\bf \ubeta} & {\bf whole} & {\bf CC} & {\bf EM} & {\bf MI} & {\bf whole} & {\bf CC} & {\bf EM} & {\bf MI} & {\bf whole} & {\bf CC} & {\bf EM} & {\bf MI} & {\bf whole} & {\bf CC} & {\bf EM} & {\bf MI} \\
			\hline
		\multicolumn{ 1}{c}{n=60} & $ y \le 1 $ & 1.000 & 1.071 & 5.598 & 2.056 & 2.699 & 0.071 & 4.598 & \textbf{1.056} & 1.699 & 0.3552 & 33.3628 & 6.1152 & \textbf{4.5660} & 0.952 & 0.326 & \textbf{0.888} & 0.389 \\
		\multicolumn{ 1}{c}{} & $ y \le 2 $ & -0.600 & -0.663 & 1.754 & -0.171 & 0.727 & 0.063 & 2.354 & \textbf{0.429} & 1.327 & 0.3302 & 11.0449 & \textbf{2.7097} & 2.9129 & 0.951 & 0.690 & \textbf{0.906} & 0.365 \\
		\multicolumn{ 1}{c}{} & $ x_1 = 1 $ & -1.000 & -1.095 & -2.850 & -1.583 & -0.898 & 0.095 & 1.850 & 0.583 & \textbf{0.102} & 0.4590 & 11.0145 & 2.9001 & \textbf{0.7048} & 0.945 & 0.959 & \textbf{0.949} & 0.919 \\
		\multicolumn{ 1}{c}{} & $ x_3 $ & 0.005 & 0.005 & 0.040 & 0.015 & 0.013 & 0.000 & 0.034 & 0.010 & \textbf{0.008} & 0.0002 & 0.0035 & 0.0011 & \textbf{0.0004} & 0.958 & 0.872 & \textbf{0.926} & 0.917 \\
		\multicolumn{ 1}{c}{} & $ x_4 $ & -0.100 & -0.110 & -0.181 & -0.140 & -0.064 & 0.010 & 0.081 & 0.040 & \textbf{0.036} & 0.0010 & 0.0180 & 0.0068 & \textbf{0.0018} & 0.951 & 0.960 & \textbf{0.940} & 0.494 \\
		&       &       &       &       &       &       &       &       &       &       &  \\
		\multicolumn{ 1}{c}{n=150} & $ y \le 1 $ & 1.000 & 1.019 & 4.298 & 1.162 & 2.510 & 0.019 & 3.298 & \textbf{0.162} & 1.510 & 0.1183 & 11.8009 & \textbf{0.9018} & 3.2673 & 0.950 & 0.004 & \textbf{0.888} & 0.206 \\
		\multicolumn{ 1}{c}{} & $ y \le 2 $ & -0.600 & -0.615 & 1.214 & -0.534 & 0.701 & 0.015 & 1.814 & \textbf{0.066} & 1.301 & 0.1069 & 3.8526 & \textbf{0.2995} & 2.3326 & 0.954 & 0.151 & \textbf{0.942} & 0.177 \\
		\multicolumn{ 1}{c}{} & $ x_1 = 1 $ & -1.000 & -1.033 & -2.047 & -1.109 & -0.805 & 0.033 & 1.047 & \textbf{0.109} & 0.195 & 0.1601 & 1.8011 & 0.4155 & \textbf{0.2591} & 0.949 & 0.786 & \textbf{0.929} & 0.866 \\
		\multicolumn{ 1}{c}{} & $ x_3 $ & 0.005 & 0.005 & 0.032 & 0.007 & 0.013 & 0.000 & 0.027 & \textbf{0.002} & 0.008 & 0.0001 & 0.0010 & \textbf{0.0002} & 0.0002 & 0.958 & 0.872 & \textbf{0.926} & 0.917 \\
		\multicolumn{ 1}{c}{} & $ x_4 $ & -0.100 & -0.103 & -0.146 & -0.109 & -0.060 & 0.003 & 0.046 & \textbf{0.009} & 0.040 & 0.0003 & 0.0032 & \textbf{0.0011} & 0.0018 & 0.956 & 0.753 & \textbf{0.938} & 0.129 \\
		&       &       &       &       &       &       &       &       &       &       &  \\
		\multicolumn{ 1}{c}{n=250} & $ y \le 1 $ & 1.000 & 1.013 & 4.177 & 1.064 & 2.508 & 0.013 & 3.177 & \textbf{0.064} & 1.508 & 0.0669 & 10.5452 & \textbf{0.3269} & 3.1373 & 0.957 & 0.000 & \textbf{0.917} & 0.146 \\
		\multicolumn{ 1}{c}{} & $ y \le 2 $ & -0.600 & -0.618 & 1.135 & -0.592 & 0.696 & 0.018 & 1.735 & \textbf{0.008} & 1.296 & 0.0630 & 3.2883 & \textbf{0.1273} & 2.2040 & 0.954 & 0.035 & \textbf{0.954} & 0.103 \\
		\multicolumn{ 1}{c}{} & $ x_1 = 1 $ & -1.000 & -1.021 & -1.957 & -1.054 & -0.806 & 0.021 & 0.957 & \textbf{0.054} & 0.194 & 0.0844 & 1.2779 & 0.1862 & \textbf{0.1859} & 0.954 & 0.628 & \textbf{0.944} & 0.814 \\
		\multicolumn{ 1}{c}{} & $ x_3 $ & 0.005 & 0.005 & 0.030 & 0.005 & 0.012 & 0.000 & 0.025 & \textbf{0.000} & 0.007 & 0.0000 & 0.0008 & \textbf{0.0001} & 0.0002 & 0.944 & 0.448 & \textbf{0.948} & 0.743 \\
		\multicolumn{ 1}{c}{} & $ x_4 $ & -0.100 & -0.103 & -0.142 & -0.105 & -0.059 & 0.003 & 0.042 & \textbf{0.005} & 0.041 & 0.0002 & 0.0023 & \textbf{0.0004} & 0.0018 & 0.939 & 0.527 & \textbf{0.930} & 0.048 \\
		&       &       &       &       &       &       &       &       &       &       &  \\
		\multicolumn{ 1}{c}{n=500} & $ y \le 1 $  & 1.000 & 1.011 & 4.095 & 1.072 & 2.468 & 0.011 & 3.095 & \textbf{0.072} & 1.468 & 0.0354 & 9.7817 & \textbf{0.1684} & 2.9521 & 0.943 & 0.000 & \textbf{0.930} & 0.086 \\
		\multicolumn{ 1}{c}{} & $ y \le 2 $  & -0.600 & -0.605 & 1.118 & -0.577 & 0.672 & 0.005 & 1.718 & \textbf{0.023} & 1.272 & 0.0318 & 3.0653 & \textbf{0.0596} & 2.0941 & 0.942 & 0.000 & \textbf{0.939} & 0.067 \\
		\multicolumn{ 1}{c}{} & $ x_1 = 1 $ & -1.000 & -1.010 & -1.921 & -1.048 & -0.786 & 0.010 & 0.921 & \textbf{0.048} & 0.214 & 0.0437 & 1.0064 & \textbf{0.0917} & 0.1333 & 0.944 & 0.324 & \textbf{0.939} & 0.717  \\
		\multicolumn{ 1}{c}{} & $ x_3 $ & 0.005 & 0.005 & 0.030 & 0.006 & 0.012 & 0.000 & 0.025 & \textbf{0.001} & 0.007 & 0.0000 & 0.0007 & \textbf{0.0000} & 0.0001 & 0.944 & 0.324 & \textbf{0.939} & 0.717 \\
		\multicolumn{ 1}{c}{} & $ x_4 $ & -0.100 & -0.101 & -0.139 & -0.103 & -0.059 & 0.001 & 0.039 & \textbf{0.003} & 0.041 & 0.0001 & 0.0018 & \textbf{0.0002} & 0.0018 & 0.957 & 0.221 & \textbf{0.942} & 0.006 \\
  &       &       &       &       &       &       &       &       &       &       &  \\
		\multicolumn{ 1}{c}{n= 1000} & $ y \le 1 $  & 1.000  & 1.000 & 3.989 & 0.999 & 2.465 & 0.000 & 2.989 & \textbf{0.001} & 1.465 & 0.0162 & 9.0196 & \textbf{0.0685} & 2.8772 & 0.957 & 0.000 & \textbf{0.932} & 0.046 \\
		\multicolumn{ 1}{c}{} & $ y \le 2 $  & -0.600 & -0.608 & 1.082 & -0.605 & 0.699 & 0.008 & 1.682 & \textbf{0.005} & 1.299 & 0.0151 & 2.8807 & \textbf{0.0255} & 2.1185 & 0.954 & 0.000 & \textbf{0.948} & 0.025 \\
		\multicolumn{ 1}{c}{} & $ x_1 = 1 $ & -1.000 & -0.999 & -1.850 & -1.001 & -0.786 & 0.001 & 0.850 & \textbf{0.001} & 0.214 & 0.0206 & 0.7945 & \textbf{0.0391} & 0.1101 & 0.948 & 0.096 & \textbf{0.957} & 0.540 \\
		\multicolumn{ 1}{c}{} & $ x_3 $ & 0.005 & 0.005 & 0.029 & 0.005 & 0.013 & 0.000 & 0.024 & \textbf{0.000} & 0.008 & 0.0000 & 0.0006 & \textbf{0.0000} & 0.0001 & 0.953 & 0.015 & \textbf{0.960} & 0.462 \\
		\multicolumn{ 1}{c}{} & $ x_4 $ & -0.100 & -0.100 & -0.136 & -0.101 & -0.058 & 0.000 & 0.036 & \textbf{0.001} & 0.042 & 0.0000 & 0.0014 & \textbf{0.0001} & 0.0018 & 0.946 & 0.032 & \textbf{0.938} & 0.000 \\
		\hline
	\end{tabular}
}
\end{table}


    
\begin{table}
  \centering
  {\scriptsize
  \caption{Results of the model of $ logit(R)$ on the regressors that include $ Y $ as a covariate.}
  \begin{tabular}{lrrrr}
  \hline
         Parameter & Estimate & Standard Error & P-value & 95\% Confidence Interval \\
    \hline
    Intercept & 0.587 & 0.896 & 0.512 & (-1.169, 2.344) \\
    $ Y $ & -1.233 & 0.535 & 0.021 & (-2.281, -0.184) \\
    Dose\_5MG & -1.776 & 0.518 & 0.001 & (-2.791, -0.762) \\
    Dose\_10MG & -2.452 & 0.600 & 0.000 & (-3.628, -1.275) \\
    AGEYR & -0.006 & 0.011 & 0.594 & (-0.027, 0.015) \\
    SEX   & -0.462 & 0.269 & 0.086 & (-0.990, 0.065) \\
    WEIGHT & 0.008 & 0.006 & 0.188 & (-0.004, 0.020) \\
    ONSETAGE & -0.011 & 0.010 & 0.903 & (-0.021, 0.019) \\
    \hline
    \end{tabular}%
  \label{table_logit_r_Y_EM}
  }
\end{table}%


    
\begin{table}
  \centering
  \caption{Results of the five point PGA response rates from the psoriasis study}
  {\scriptsize
    \begin{tabular}{lrr|rr|rr}
          & \multicolumn{2}{c}{CC-Estimates} & \multicolumn{2}{c}{EM-Estimates} & \multicolumn{2}{c}{MI-Estimates} \\
    \hline
    Parameter & \multicolumn{1}{r}{Odds} & \multicolumn{1}{r}{95\% Confidence} & \multicolumn{1}{r}{Odds} & \multicolumn{1}{r}{95\% Confidence} & \multicolumn{1}{r}{Odds} & \multicolumn{1}{r}{95\% Confidence} \\
          & \multicolumn{1}{r}{Ratio} & \multicolumn{1}{l}{Interval} & \multicolumn{1}{r}{Ratio} & \multicolumn{1}{l}{Interval} & \multicolumn{1}{r}{Ratio} & \multicolumn{1}{l}{Interval} \\
    \hline
    Dose\_5MG & 4.683 & (3.278, 6.696) & 3.714 & (2.494, 5.528) & 3.408 & (2.471, 4.699) \\
    Dose\_10MG & 8.998 & (6.237, 12.981) & 6.436 & (4.133, 10.024) & 6.540 & (4.705, 9.091) \\
    AGEYR & 1.007 & (0.995, 1.019) & 1.006 & (0.994, 1.018) & 1.007 & (0.997, 1.017) \\
    SEX   & 1.279 & (0.974, 1.679) & 1.201 & (0.918, 1.571) & 1.224 & (0.948, 1.579) \\
    WEIGHT & 0.992 & (0.986, 0.998) & 0.992 & (0.986, 0.998) & 0.993 & (0.987, 0.999) \\
    ONSETAGE & 0.989 & (0.977, 1.001) & 1.012 & (1.000, 1.024) & 1.005 & (0.995, 1.015) \\
    \hline
    \end{tabular}%
    }
  \label{tab:table_tofareslt}%
\end{table}%


\appendix

\section{The Computer Program}

We provide the SAS MACRO program that was used to carry out the data analysis in Section \ref{section_real_data} at Data Files and at GitHUB: 
\newline
https://github.com/arnabkrmaity/ProportionalOddsMissingResponse

\end{document}


\maketitle

\section{Visual Summary of Simulation Outputs}

\begin{figure}[ht]
	\centering
	\includegraphics[width = \textwidth, height = 10 cm]{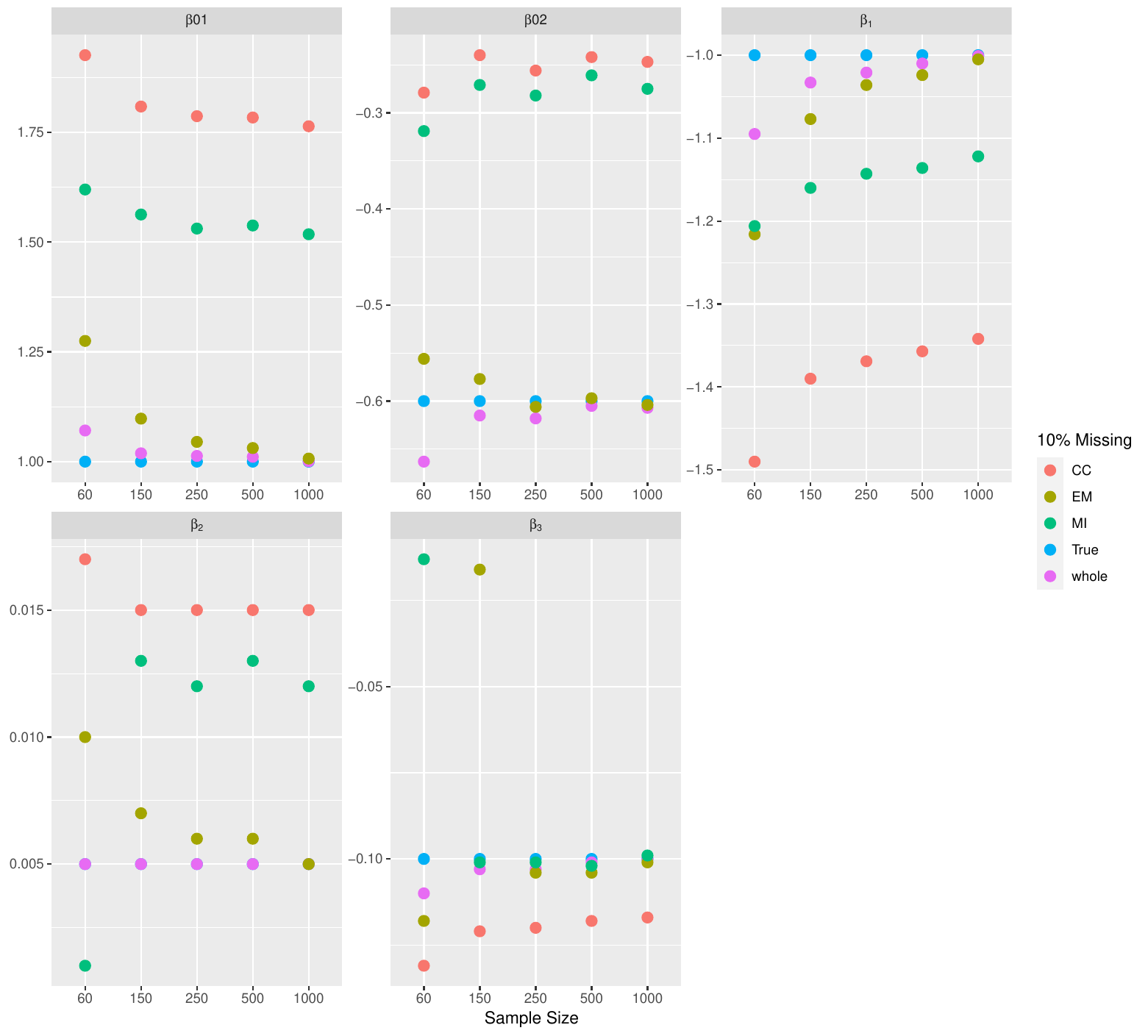}
	\caption{The mean estimates of $ \beta $ from 1000 replications using different methods along with the true values when about 10\% data are missing.}
 \label{figure_simulation_missing_10}
\end{figure}

\begin{figure}[ht]
	\centering
	\includegraphics[width = \textwidth, height = 10 cm]{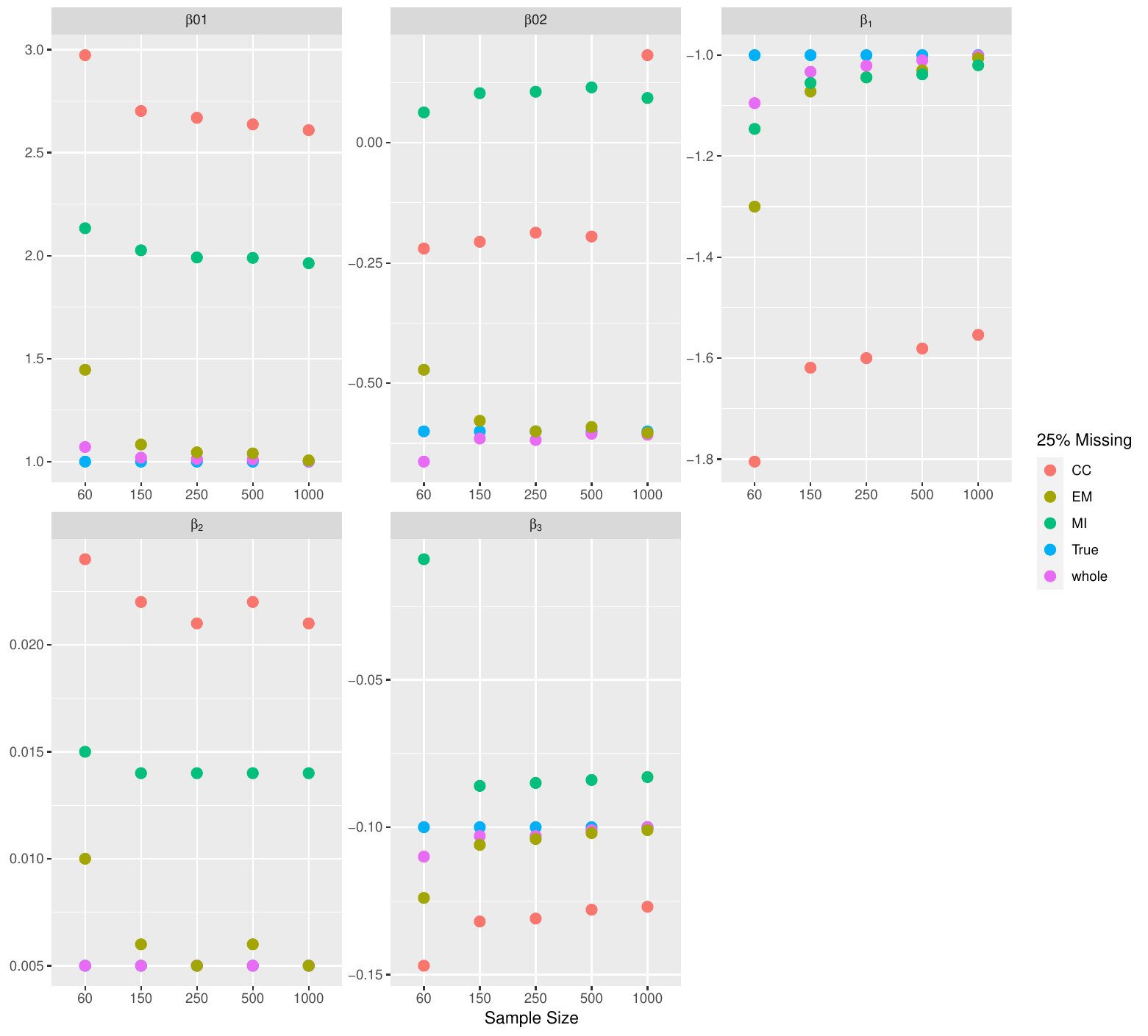}
	\caption{The mean estimates of $ \beta $ from 1000 replications using different methods along with the true values when about 25\% data are missing.}
 \label{figure_simulation_missing_25}
\end{figure}

\begin{figure}[ht]
	\centering
	\includegraphics[width = \textwidth, height = 10 cm]{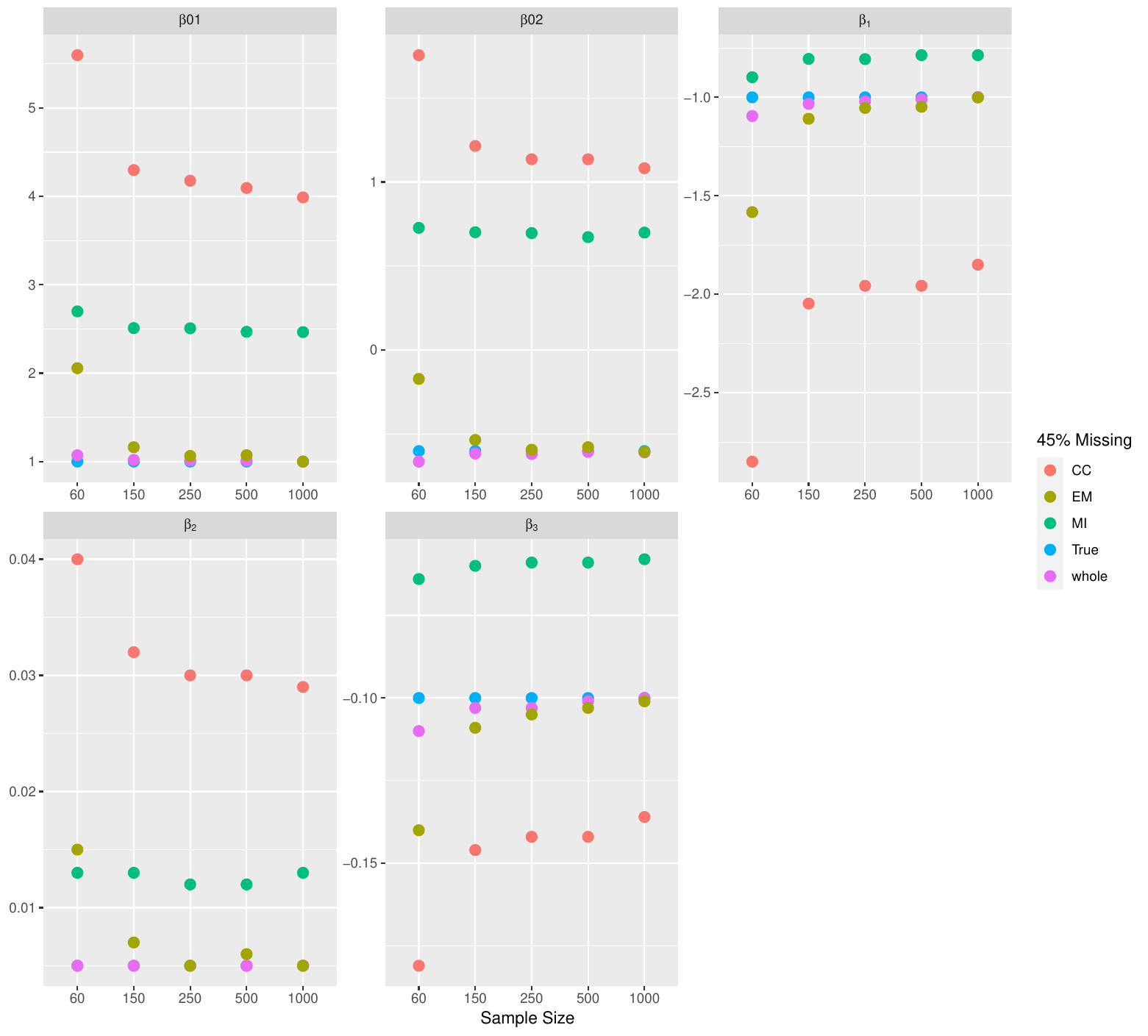}
	\caption{The mean estimates of $ \beta $ from 1000 replications using different methods along with the true values when about 45\% data are missing.}
 \label{figure_simulation_missing_45}
\end{figure}

We present simulation results using different methods in Figure \ref{figure_simulation_missing_10} when about 10\% responses are missing; the results are presented in Figure \ref{figure_simulation_missing_25} when there are about 25\% missing responses; finally, we present the simulations results in Figure \ref{figure_simulation_missing_45} when there exists 45\% missing responses in the simulated data.

\section{Additional Simulation}  \label{section_simulation_supp}

\subsection{Operating Characteristics with 5 Response Categories in Simulated Examples}

This Section provides outputs when we generated the simulated datasets with 5 categories in the response variable to assess the performance of the proposed method. 

We generate 100 simulated datasets keeping the true values of $$ \beta = (0.6, 0.5, -0.2, -0.7, -1.3, 0.008, -0.02)^{T} $$ to summarize the different matrices when about 25\% observations are missing. We present the simulations summaries in Tables \ref{table_sim1_mis10_cat5} and \ref{table_sim1_mis10_cat5_CP}. We note that, the results corroborate the evidence which was obtained in the examples having 3 categories in the response data, that is, the proposed expectation maximization (EM) method is the second best method (after ``whole") in reducing the bias outperforming the complete case (CC) method.

\begin{table}[h!]
	{\tiny
		\centering
		\caption{Simulation with $\sim$ 25\% missing responses and 5 categories in the response: Simulation results for $n=60, 150,250,500$ with $1000$ replications for each scenario. Summary statistics (1) $E[\widehat{\beta}]$ is the mean of $1000$ estimators (2) Absolute bias = $| E[\hat{\beta}]-\beta |$ (3) MSE = bias$^2$+SD$^2$. }
  \label{table_sim1_mis10_cat5}
		\begin{tabular}{rrrrrrrrrrrrrrr}
			\hline
			&       & {\bf } & \multicolumn{4}{|c}{{\bf $E[\widehat{\beta}]$}} & \multicolumn{ 4}{|c}{{\bf Absolute Bias}} & \multicolumn{ 4}{|c}{{\bf MSE}} \\
			&       & {\bf \ubeta} & {\bf whole} & {\bf CC} & {\bf EM} & {\bf MI} & {\bf whole} & {\bf CC} & {\bf EM} & {\bf MI} & {\bf whole} & {\bf CC} & {\bf EM} & {\bf MI} \\
			\hline
			\multicolumn{ 1}{c}{n=60} & $ y \le 1 $ & 0.600 & 0.658 & 1.724 & 1.304 & 1.219 & 0.058 & 1.124 & 0.704 & 0.619 & 0.2442 & 1.7381 & 1.2219 & 0.7427 \\
			\multicolumn{ 1}{c}{}     & $ y \le 2 $ & 0.500 & 0.512 & 1.535 & 1.143 & 1.047 & 0.012 & 1.035 & 0.643 & 0.547 & 0.2337 & 1.5294 & 1.1247 & 0.6470 \\
			\multicolumn{ 1}{c}{}     & $ y \le 3 $ & -0.200 & -0.216 & 0.633 & 0.381 & 0.225 & 0.016 & 0.833 & 0.582 & 0.425 & 0.2228 & 1.1047 & 1.0045 & 0.5074 \\
			\multicolumn{ 1}{c}{}     & $ y \le 4 $ & -0.700 & -0.760 & -0.022 & -0.185 & -0.365 & 0.060 & 0.678 & 0.515 & 0.335 & 0.2375 & 0.8690 & 0.9555 & 0.4713 \\
			\multicolumn{ 1}{c}{}     & $ x_1 = 1 $ & -1.300 & -1.394 & -2.034 & -1.691 & -1.412 & 0.094 & 0.734 & 0.391 & 0.112 & 0.3619 & 1.1417 & 0.7509 & 0.3404 \\
			\multicolumn{ 1}{c}{}     & $ x_3 $     & 0.008 & 0.009 & 0.010 & 0.010 & 0.008 & 0.001 & 0.002 & 0.002 & 0.000 & 0.0002 & 0.0003 & 0.0003 & 0.0002 \\
			\multicolumn{ 1}{c}{}     & $ x_4 $      & -0.020 & -0.023 & -0.024 & -0.023 & -0.017 & 0.003 & 0.004 & 0.003 & 0.003 & 0.0003 & 0.0005 & 0.0004 & 0.0003 \\
			&       &       &       &       &       &       &       &       &       &       &      &      &    
			\\
			\multicolumn{ 1}{c}{n= 150} & $ y \le 1 $ & 0.600 & 0.602 & 1.571 & 0.972 & 1.105 & 0.002 & 0.971 & 0.372 & 0.505 & 0.0839 & 1.0701 & 0.5121 & 0.4453 \\
			\multicolumn{ 1}{c}{}     & $ y \le 2 $ & 0.500 & 0.498 & 1.450 & 0.867 & 0.994 & 0.002 & 0.950 & 0.367 & 0.494 & 0.0827 & 1.0241 & 0.4969 & 0.4280 \\
			\multicolumn{ 1}{c}{}     & $ y \le 3 $ & -0.200 & -0.218 & 0.577 & 0.131 & 0.198 & 0.018 & 0.777 & 0.331 & 0.398 & 0.0801 & 0.7177 & 0.4380 & 0.3366 \\
			\multicolumn{ 1}{c}{}     & $ y \le 4 $ & -0.700 & -0.737 & -0.038 & -0.410 & -0.376 & 0.037 & 0.662 & 0.290 & 0.324 & 0.0872 & 0.5529 & 0.3975 & 0.2875 \\
			\multicolumn{ 1}{c}{}     & $ x_1 = 1 $ & -1.300 & -1.325 & -1.897 & -1.497 & -1.327 & 0.025 & 0.597 & 0.197 & 0.027 & 0.1233 & 0.5262 & 0.2361 & 0.1168 \\
			\multicolumn{ 1}{c}{}     & $ x_3 $      & 0.008 & 0.009 & 0.010 & 0.009 & 0.008 & 0.001 & 0.002 & 0.001 & 0.000 & 0.0001 & 0.0001 & 0.0001 & 0.0001 \\
			\multicolumn{ 1}{c}{}     & $ x_4 $      & -0.020 & -0.021 & -0.022 & -0.021 & -0.016 & 0.001 & 0.002 & 0.001 & 0.004 & 0.0001 & 0.0002 & 0.0001 & 0.0001 \\
			&       &       &       &       &       &       &       &       &       &       &      &      &    \\
			\multicolumn{ 1}{c}{n=250} & $ y \le 1 $  & 0.600 & 0.608 & 1.549 & 0.847 & 1.124 & 0.008 & 0.949 & 0.247 & 0.524 & 0.0475 & 0.9685 & 0.3116 & 0.4016 \\
		     & $ y \le 2 $ & 0.500 & 0.509 & 1.438 & 0.750 & 1.021 & 0.009 & 0.938 & 0.250 & 0.521 & 0.0474 & 0.9469 & 0.3072 & 0.3972 \\
		     & $ y \le 3 $ & -0.200 & -0.201 & 0.573 & 0.019 & 0.226 & 0.001 & 0.773 & 0.219 & 0.426 & 0.0442 & 0.6579 & 0.2475 & 0.3047 \\
		     & $ y \le 4 $ & -0.700 & -0.708 & -0.024 & -0.502 & -0.336 & 0.008 & 0.676 & 0.198 & 0.364 & 0.0473 & 0.5164 & 0.2299 & 0.2628 \\
		     & $ x_1 = 1 $   & -1.300 & -1.317 & -1.863 & -1.444 & -1.322 & 0.017 & 0.563 & 0.144 & 0.022 & 0.0724 & 0.4106 & 0.1457 & 0.0709 \\
		     & $ x_3 $     & 0.008 & 0.008 & 0.010 & 0.008 & 0.007 & 0.000 & 0.002 & 0.000 & 0.001 & 0.0000 & 0.0001 & 0.0000 & 0.0000 \\
		     & $ x_4 $     & -0.020 & -0.021 & -0.022 & -0.021 & -0.016 & 0.001 & 0.002 & 0.001 & 0.004 & 0.0001 & 0.0001 & 0.0001 & 0.0001 \\
			&       &       &       &       &       &       &       &       &       &       &  \\
			\multicolumn{ 1}{c}{n=500} & $ y \le 1 $ & 0.600 & 0.606 & 1.529 & 0.708 & 1.096 & 0.006 & 0.929 & 0.108 & 0.496 & 0.0250 & 0.8972 & 0.1261 & 0.3546 \\
		     & $ y \le 2 $ & 0.500 & 0.505 & 1.417 & 0.609 & 0.994 & 0.005 & 0.917 & 0.109 & 0.494 & 0.0244 & 0.8735 & 0.1209 & 0.3481 \\
		     & $ y \le 3 $ & -0.200 & -0.200 & 0.561 & -0.107 & 0.209 & 0.000 & 0.761 & 0.093 & 0.409 & 0.0227 & 0.6077 & 0.0922 & 0.2670 \\
		     & $ y \le 4 $ & -0.700 & -0.706 & -0.033 & -0.622 & -0.356 & 0.006 & 0.667 & 0.078 & 0.344 & 0.0244 & 0.4738 & 0.0795 & 0.2277 \\
		     & $ x_1 = 1 $   & -1.300 & -1.311 & -1.844 & -1.376 & -1.322 & 0.011 & 0.544 & 0.076 & 0.022 & 0.0375 & 0.3435 & 0.0759 & 0.0384 \\
		     & $ x_3 $     & 0.008 & 0.008 & 0.009 & 0.008 & 0.007 & 0.000 & 0.001 & 0.000 & 0.001 & 0.0000 & 0.0000 & 0.0000 & 0.0000 \\
		     & $ x_4 $     & -0.020 & -0.020 & -0.021 & -0.020 & -0.015 & 0.000 & 0.001 & 0.000 & 0.005 & 0.0000 & 0.0000 & 0.0000 & 0.0000 \\
       	&       &       &       &       &       &       &       &       &       &       &  \\
			\multicolumn{ 1}{c}{n= 1000} & $ y \le 1 $ & 0.600 & 0.600 & 1.523 & 0.657 & 1.122 & 0.000 & 0.923 & 0.057 & 0.522 & 0.012 & 0.868 & 0.049 & 0.3634 \\
		     & $ y \le 2 $ & 0.500 & 0.500 & 1.411 & 0.557 & 1.022 & 0.000 & 0.911 & 0.057 & 0.522 & 0.011 & 0.846 & 0.046 & 0.3642 \\
		     & $ y \le 3 $ & -0.200 & -0.202 & 0.558 & -0.154 & 0.228 & 0.002 & 0.758 & 0.046 & 0.428 & 0.011 & 0.590 & 0.032 & 0.2670 \\
		     & $ y \le 4 $ & -0.700 & -0.705 & -0.032 & -0.663 & -0.318 & 0.005 & 0.668 & 0.037 & 0.382 & 0.012 & 0.461 & 0.027 & 0.2400  \\
		     & $ x_1 = 1 $   & -1.300 & -1.306 & -1.836 & -1.346 & -1.317 & 0.006 & 0.536 & 0.046 & 0.017 & 0.017 & 0.311 & 0.035 & 0.020 \\
		     & $ x_3 $     & 0.008 & 0.008 & 0.010 & 0.008 & 0.007 & 0.000 & 0.002 & 0.000 & 0.001 & 0.000 & 0.000 & 0.000 & 0.0000 \\
		     & $ x_4 $     & -0.020 & -0.020 & -0.021 & -0.020 & -0.015 & 0.000 & 0.001 & 0.000 & 0.005 & 0.000 & 0.000 & 0.000 & 0.0000 \\
		\end{tabular}
		
	}
\end{table}

\begin{table}
\centering
{\tiny
	\caption{Simulation with $\sim$ 25\% missing responses and 5 categories in the response: Results for $n=60, 150,250,500$ with $1000$ replications. Summary statistics (1)SD is the standard deviation of $1000$ $\widehat{\beta}$ (2) is the mean of $1000$ SE estimates (3) 95\% CP is the 95\% coverage percentage. }
 \label{table_sim1_mis10_cat5_CP}
	\begin{tabular}{rrrrrrrrrrrrrr}
		\hline
		{\bf } & {\bf } & \multicolumn{ 4}{c}{{\bf SD}} & \multicolumn{ 4}{c}{{\bf $E[\text{SE}]$}} & \multicolumn{ 4}{c}{{\bf 95\%CP}} \\
		{\bf } & {\bf } & {\bf whole} & {\bf CC} & {\bf EM} & {\bf MI} & {\bf whole} & {\bf CC} & {\bf EM} & {\bf MI} & {\bf whole} & {\bf CC} & {\bf EM} & {\bf MI} \\
		\hline
		n=60 & $ y \le 1 $ & 0.4907 & 0.6885 & 0.8520 & 0.5999 & 0.4630 & 0.6468 & 0.7343 & 0.4780 & 0.9395 & 0.5937 & 0.7537 & 0.7039 \\
		     & $ y \le 2 $ & 0.4833 & 0.6771 & 0.8434 & 0.5898 & 0.4596 & 0.6359 & 0.7241 & 0.4712 & 0.9476 & 0.6435 & 0.7671 & 0.7402 \\
		     & $ y \le 3 $ & 0.4718 & 0.6407 & 0.8166 & 0.5715 & 0.4534 & 0.5927 & 0.6834 & 0.4499 & 0.9530 & 0.7115 & 0.7704 & 0.7840 \\
		     & $ y \le 4 $ & 0.4837 & 0.6396 & 0.8308 & 0.5992 & 0.4683 & 0.5837 & 0.6766 & 0.4553 & 0.9476 & 0.7825 & 0.7787 & 0.8097 \\
		     & $ x_1 = 1 $  & 0.5942 & 0.7764 & 0.7733 & 0.5725 & 0.5669 & 0.7222 & 0.7513 & 0.5539 & 0.9476 & 0.8459 & 0.9301 & 0.9456 \\
		     & $ x_2 $     & 0.0139 & 0.0167 & 0.0161 & 0.0138 & 0.0135 & 0.0154 & 0.0148 & 0.0129 & 0.9570 & 0.9532 & 0.9551 & 0.9486 \\
		     & $ x_3 $      & 0.0183 & 0.0209 & 0.0206 & 0.0164 & 0.0174 & 0.0197 & 0.0189 & 0.0155 & 0.9664 & 0.9562 & 0.9418 & 0.9411 \\
		&       &       &       &       &       &       &       &       &       &  \\
		n= 150 & $ y\le 1 $ & 0.2896 & 0.3561 & 0.6111 & 0.4363 & 0.2830 & 0.3698 & 0.4830 & 0.2906 & 0.9593 & 0.2287 & 0.7837 & 0.5558 \\
		     & $ y \le 2 $ & 0.2875 & 0.3493 & 0.6016 & 0.4288 & 0.2815 & 0.3654 & 0.4758 & 0.2879 & 0.9614 & 0.2352 & 0.7872 & 0.5580 \\
		     & $ y \le 3 $ & 0.2825 & 0.3376 & 0.5731 & 0.4222 & 0.2782 & 0.3398 & 0.4323 & 0.2754 & 0.9531 & 0.3654 & 0.7953 & 0.6127 \\
		     & $ y \le 4 $ & 0.2930 & 0.3389 & 0.5596 & 0.4269 & 0.2875 & 0.3359 & 0.4223 & 0.2787 & 0.9489 & 0.4912 & 0.8209 & 0.6586 \\
		     & $ x_1 = 1 $   & 0.3503 & 0.4122 & 0.4440 & 0.3407 & 0.3467 & 0.4161 & 0.4551 & 0.3380 & 0.9489 & 0.7298 & 0.9360 & 0.9464 \\
		     & $ x_2 $      & 0.0085 & 0.0097 & 0.0092 & 0.0078 & 0.0081 & 0.0092 & 0.0088 & 0.0077 & 0.9468 & 0.9387 & 0.9535 & 0.9650 \\
		     & $ x_3 $      & 0.0107 & 0.0123 & 0.0121 & 0.0097 & 0.0103 & 0.0115 & 0.0112 & 0.0092 & 0.9531 & 0.9464 & 0.9384 & 0.8950 \\
		&       &       &       &       &       &       &       &       &       &  \\
		n= 250 & $ y \le 1 $ & 0.2177 & 0.2591 & 0.5005 & 0.3572 & 0.2165 & 0.2808 & 0.3977 & 0.2224 & 0.9467 & 0.0481 & 0.8562 & 0.4012 \\
		     & $ y \le 2 $ & 0.2175 & 0.2574 & 0.4945 & 0.3548 & 0.2154 & 0.2777 & 0.3906 & 0.2204 & 0.9417 & 0.0420 & 0.8617 & 0.4074 \\
		     & $ y\le 3 $ & 0.2103 & 0.2446 & 0.4468 & 0.3516 & 0.2125 & 0.2579 & 0.3442 & 0.2102 & 0.9558 & 0.1167 & 0.8761 & 0.4872 \\
		     & $ y \le 4 $ & 0.2174 & 0.2423 & 0.4368 & 0.3609 & 0.2192 & 0.2549 & 0.3305 & 0.2123 & 0.9558 & 0.2252 & 0.8894 & 0.5568 \\
		     & $ x_1 = 1 $  & 0.2685 & 0.3053 & 0.3533 & 0.2653 & 0.2655 & 0.3165 & 0.3615 & 0.2587 & 0.9477 & 0.5752 & 0.9369 & 0.9488 \\
		     & $ x_2 $   & 0.0062 & 0.0070 & 0.0067 & 0.0060 & 0.0062 & 0.0070 & 0.0067 & 0.0059 & 0.9528 & 0.9427 & 0.9502 & 0.9488 \\
		     & $ x_3 $  & 0.0080 & 0.0091 & 0.0089 & 0.0072 & 0.0078 & 0.0087 & 0.0085 & 0.0069 & 0.9467 & 0.9478 & 0.9403 & 0.8680 \\
		&       &       &       &       &       &       &       &       &       &  \\
			n= 500 & $ y \le 1 $ & 0.1581 & 0.1846 & 0.3383 & 0.3287 & 0.1526 & 0.1968 & 0.2954 & 0.1566 & 0.9460 & 0.0020 & 0.9293 & 0.2763 \\
		     & $ y \le 2 $ & 0.1562 & 0.1814 & 0.3301 & 0.3231 & 0.1518 & 0.1946 & 0.2876 & 0.1553 & 0.9480 & 0.0020 & 0.9358 & 0.2893 \\
		     & $ y \le 3 $ & 0.1508 & 0.1682 & 0.2889 & 0.3164 & 0.1498 & 0.1808 & 0.2418 & 0.1483 & 0.9470 & 0.0050 & 0.9456 & 0.3594 \\
		     & $ y \le 4 $ & 0.1559 & 0.1706 & 0.2707 & 0.3307 & 0.1546 & 0.1789 & 0.2276 & 0.1500 & 0.9460 & 0.0310 & 0.9499 & 0.4444 \\
		     & $ x_1 = 1 $   & 0.1933 & 0.2175 & 0.2646 & 0.1946 & 0.1869 & 0.2217 & 0.2625 & 0.1824 & 0.9520 & 0.3073 & 0.9423 & 0.9309 \\
		     & $ x_2 $     & 0.0044 & 0.0050 & 0.0048 & 0.0041 & 0.0043 & 0.0049 & 0.0046 & 0.0042 & 0.9400 & 0.9359 & 0.9412 & 0.9469 \\
		     & $ x_3 $    & 0.0056 & 0.0062 & 0.0059 & 0.0053 & 0.0055 & 0.0060 & 0.0058 & 0.0049 & 0.9490 & 0.9489 & 0.9467 & 0.7928 \\
       &       &       &       &       &       &       &       &       &       &  \\
			n= 1000 & $ y \le 1 $ & 0.1079 & 0.1288 & 0.2129 & 0.3012 & 0.1076 & 0.1387 & 0.2085 & 0.1107 & 0.9510 & 0.0000 & 0.9515 & 0.1500 \\
		     & $ y \le 2 $ & 0.1072 & 0.1274 & 0.2068 & 0.3021 & 0.1070 & 0.1371 & 0.2018 & 0.1097 & 0.9490 & 0.0000 & 0.9578 & 0.1580 \\
		     & $ y \le 3 $ & 0.1070 & 0.1224 & 0.1715 & 0.2902  & 0.1056 & 0.1274 & 0.1642 & 0.1045 & 0.9440 & 0.0000 & 0.9599 & 0.2280 \\
		     & $ y \le 4 $ & 0.1108 & 0.1234 & 0.1610 & 0.3061 & 0.1090 & 0.1261 & 0.1530 & 0.1055 & 0.9520 & 0.0000 & 0.9652 & 0.2770 \\
		     & $ x_1 = 1 $ & 0.1312 & 0.1521 & 0.1804 & 0.1412 & 0.1318 & 0.1562 & 0.1857 & 0.1285 & 0.9460 & 0.0510 & 0.9536 & 0.9250 \\
		     & $ x_2 $     & 0.0031 & 0.0034 & 0.0032 & 0.0029 & 0.0031 & 0.0034 & 0.0032 & 0.0029 & 0.9490 & 0.9270 & 0.9462 & 0.9380 \\
		     & $ x_3 $     & 0.0037 & 0.0041 & 0.0039 & 0.0038 & 0.0038 & 0.0042 & 0.0040 & 0.0034 & 0.9610 & 0.9650 & 0.9599 & 0.6180 \\
		\hline
	\end{tabular}
}
\end{table}

\subsection{Operating Characteristics with different combination of $\alpha$ and $\beta$}

Here we have further explored the impact of missingness with other combinations of $ \alpha $ and $ \beta $ parameters through simulations and the findings remains the same. For example, in the following in Table \ref{table_sim1mis1001}, we present a small simulation study to summarize the performance of the proposed method as compared to the existing methods. We set $ \alpha = (4.8, 1, -0.6, 0.05, -0.1, -3)^T $ and $ \beta = (1, -0.6, 0.5, -0.05, 0.1)^T $. The sample size was set to 250, and 45\% of data are missing. It can be noticed that, the bias is less when the proposed EM algorithm is fitted to the data than that of other methods.

\begin{table}[ht]

		\centering
		\caption{Simulation with different true values of the parameters: Simulation results for $n= 250$ with $1000$ replications. Summary statistics (1) $E[\widehat{\beta}]$ is the mean of $1000$ estimators (2) Absolute bias = $| E[\hat{\beta}]-\beta |$ (4) MSE = bias$^2$+SD$^2$ (4) 95\% CP is the 95\% coverage percentage. The best values among ``CC", ``EM",and ``MI"are marked in bold. \label{table_sim1mis1001}}
\resizebox{\textwidth}{!}{\begin{minipage}{1.9\textwidth}

\begin{tabular}{rrrrrrrrrrrrrrrrrrr}
			\hline
			&       & {\bf } & \multicolumn{4}{|c}{{\bf $E[\widehat{\beta}]$}} & \multicolumn{ 4}{|c}{{\bf Absolute Bias}} & \multicolumn{ 4}{|c}{{\bf MSE}} & \multicolumn{ 4}{|c}{{\bf 95\% CP}} \\
			&       & {\bf \ubeta} & {\bf whole} & {\bf CC} & {\bf EM} & {\bf MI} & {\bf whole} & {\bf CC} & {\bf EM} & {\bf MI} & {\bf whole} & {\bf CC} & {\bf EM} & {\bf MI} & {\bf whole} & {\bf CC} & {\bf EM} & {\bf MI} \\
			\hline
			\multicolumn{ 1}{c}{n=250} & $ y \ge 1 $ & 1.000 & 1.022 & 3.862 & \textbf{1.570} & 4.651 & 0.022 & 2.862 & \textbf{0.570} & 3.651 & 0.0800 & 11.1198 & \textbf{2.3764} & 15.7143 & 0.910 & 0.238 & \textbf{0.940} & 0.095 \\
			\multicolumn{ 1}{c}{} & $ y \ge 2 $ & -0.600 & -0.607 & 1.773 & \textbf{-0.285} & 2.486 & 0.007 & 2.373 & \textbf{0.315} & 3.086 & 0.0730 & 8.6316 & \textbf{1.0780} & 10.5876 & 0.940 & 0.048 & \textbf{0.920} & 0.000 \\
			\multicolumn{ 1}{c}{} & $ x_1 = 1 $ & 0.500 & 0.498 & 0.628 & \textbf{0.549} & 0.398 & 0.002 & 0.128 & \textbf{0.049} & 0.102 & 0.0967 & 3.6833 & \textbf{0.2277} & 1.1265 & 0.940 & \textbf{0.952} & \textbf{0.952} & 0.920 \\
			\multicolumn{ 1}{c}{} & $ x_3 $   & -0.050 & -0.052 & -0.022 & \textbf{-0.055} & 0.010 & 0.002 & 0.028 & \textbf{0.005} & 0.059 & 0.0001 & 0.0018 & \textbf{0.0009} & 0.0043 & 0.970 & 0.762 & \textbf{0.960} & 0.048 \\
			\multicolumn{ 1}{c}{} & $ x_4 $   & 0.100 & 0.101 & 0.049 & \textbf{0.112} & 0.007 & 0.001 & 0.051 & \textbf{0.012} & 0.093 & 0.0001 & 0.0039 & \textbf{0.0016} & 0.0091 & 0.950 & 0.429 & \textbf{0.980} & 0.000 \\
			\hline
		\end{tabular}
		
	\end{minipage} }
\end{table}

\subsection{Firth Correction}

In this Section, we have carried out a small simulation by incorporating Firth \citep{firth1993bias} correction. The results of the estimates are given below in column ``Firth" in Table \ref{table_firth}. Please note that the bias of the estimates corresponding to Firth type correction is less than the same using EM estimates given in the paper

\begin{table}[ht]
	{\tiny
		\centering
		\caption{Simulation with $\sim$ 10\% missing non-responses and with Firth correction incorporated: Simulation results for $n=60, 150,250,500$ with $1000$ replications for each scenario. Summary statistics (1) $E[\widehat{\beta}]$ is the mean of $1000$ estimators (2) Absolute bias = $| E[\hat{\beta}]-\beta |$. }
  \label{table_firth}
		\begin{tabular}{rrrrrrrrrrrrr}
			\hline
			&       & {\bf } & \multicolumn{5}{|c}{{\bf $E[\widehat{\beta}]$}} & \multicolumn{ 5}{|c}{{\bf Absolute Bias}} \\
			&       & {\bf \ubeta} & {\bf whole} & {\bf CC} & {\bf EM} & {\bf Firth} & {\bf MI} & {\bf whole} & {\bf CC} & {\bf EM} & {\bf Firth} & {\bf MI} \\
			\hline
			\multicolumn{ 1}{c}{n=60} & $ y \ge 1 $ & 1.000 & 1.071 & 1.926 & 1.275 & 1.222 & 1.620 & 0.071 & 0.926 & 0.275 & 0.222 & 0.626 \\
			\multicolumn{ 1}{c}{} & $ y \ge 2 $ & -0.600 & -0.663 & -0.279 & -0.556 & -0.598 & -0.319 & 0.063 & 0.321 & 0.044 & 0.002 & 0.281 \\
			\multicolumn{ 1}{c}{} & $ x_1 = 1 $ & -1.000 & -1.095 & -1.490 & -1.216 & -1.336 & -1.136 & 0.095 & 0.490 & 0.216 & 0.136 & 0.206 \\
			\multicolumn{ 1}{c}{} & $ x_3 $ & 0.005 & 0.005 & 0.017 & 0.010 & 0.015 & 0.013 & 0.000 & 0.012 & 0.005 & 0.010 & 0.008  \\
			\multicolumn{ 1}{c}{} & $ x_4 $ & -0.100 & -0.110 & -0.131 & 0.118 & -0.108 & -0.108 & -0.100 & 0.010 & 0.018 & 0.008 & 0.008  \\
			\hline
		\end{tabular}
		
	}
\end{table}

\bibliographystyle{apalike} 
\bibliography{referencePOmissing.bib}

\appendix